# Interplay between magnetic and lattice excitations and emergent multiple phase transitions in MnPSe$_{3-x}$S$_x$

Deepu Kumar[1], Nguyen The Hoang[1], Yumin Sim[1], Youngsu Choi[2], Kalaivanan Raju[3], Rajesh Kumar Ulaganathan[3], Raman Sankar[3], Maeng-Je Seong[1]*, and Kwang-Yong Choi[2]*

[1]*Department of Physics and Center for Berry Curvature-based New Phenomena (BeCaP) Chung-Ang University, Seoul 06974, Republic of Korea*
[2]*Department of Physics, Sungkyunkwan University, Suwon 16419, Republic of Korea*
[3]*Institute of Physics, Academia Sinica, Taipei 10617, Taiwan*

**Abstract**

The intricate interplay between spin and lattice degrees of freedom in two-dimensional magnetic materials plays a pivotal role in modifying their magnetic characteristics, engendering hybrid quasiparticles, and implementing functional devices. Herein, we present our comprehensive and in-depth investigations on magnetic and lattice excitations of MnPSe$_{3-x}$S$_x$ ($x$ = 0, 0.5, and 1.5) alloys, utilizing temperature- and polarization-dependent Raman scattering. Our experimental results reveal the occurrence of multiple phase transitions, evidenced by notable changes in phonon self-energy and the appearance or splitting of phonon modes. These emergent phases are tied to the development of long and short-range spin-spin correlations, as well as to spin reorientations or magnetic instabilities. Our analysis of two-magnon excitations as a function of temperature and composition showcases their hybridization with phonons whose degree weakens with increasing $x$. Moreover, the suppression of spin-dependent phonon intensity in chemically most-disordered MnPSe$_{3-x}$S$_x$ ($x$ = 1.5) suggests that chalcogen substitution offers a control knob of tuning spin and phonon dynamics by modulating concurrently superexchange pathways and a degree of trigonal distortions.

* Email: mseong@cau.ac.kr (M. J. Seong), choisky99@skku.edu (K.Y. Choi)

# 1. Introduction

The quest for two-dimensional (2D) magnetism in van der Waals (vdW) materials constitutes an exciting frontier in condensed-matter physics, especially in the context of the Mermin-Wagner theorem, which negates the possibility of long-range ordering at finite temperatures in 2D systems with isotropic interactions [1]. However, the discovery of long-range magnetic order in vdW materials such as $CrI_3$ [2] and $Cr_2Ge_2Te_6$ [3] has expanded the purview of low-dimensional magnetism and catalyzed extensive research into magnetism at the 2D limit, subsequently enlarging the family of 2D magnetic materials [4,5].

Among these materials, transition metal phosphorous trichalcogenides $MPX_3$ (M=Mn, Fe, Ni V Co; X=S, Se), which form vdW antiferromagnets (AFM), have emerged as exceptional platforms for exploring the exotic and topological properties of 2D magnetism [6–11]. Singularly, the magnetic exchange interactions, spin ordering patterns, spin dimensionality, and magnetic anisotropies are strongly dependent on the specific choice of M or X atoms. For example, $NiPS_3$ is characterized by an XXZ-type AFM ($T_N$ ~155 K) with zigzag-ordered moments lying in the *ab* plane, while $FePS_3$ embodies an Ising AFM with an out-of-plane zigzag order at $T_N$~123 K [10,12]. The Heisenberg AFM $MnPS_3$ undergoes a transition to Néel-type ordering at $T_N$ ~78 K. Conversely, its selenium counterpart $MnPSe_3$ exhibits large XY anisotropy with a similar ordering temperature of $T_N$~74 K [13]. Additionally, the spin ordering orientation relies on the species of non-magnetic ions; in $MnPSe_3$, the spins are aligned within the *ab* plane, whereas in $MnPS_3$, the ordered magnetic moments lie perpendicular to the *ab* plane. The distinct magnetic anisotropies and interactions observed in $MnPX_3$ are discussed in terms of the combined effects of trigonal distortion of the $MnX_6$ octahedra and spin-orbit coupling [13,14], highlighting the turnability of magnetic properties through chemical substitution.

In this vein, the variation among magnetic or nonmagnetic atoms can lead to the emergence of exotic magnetic phenomena in alloyed compounds, *MM′*PX$_3$ (M and M′ represent different transition metal atoms) or MP(*XX′*)$_3$ (X and X′ are different chalcogen atoms). This is owed to the interplay between competing magnetic interactions of heterogeneous magnetic ions, varying strengths of spin-orbit coupling between magnetic and chalcogen atoms, and magnetic anisotropy. A wide spectrum of studies focusing on alloying at the M site, including Ni$_{1-x}$Fe$_x$PS$_3$, Ni$_{1-x}$Mn$_x$PS$_3$, Mn$_{1-x}$Fe$_x$PSe$_3$, Mn$_x$Zn$_{1-x}$PS$_3$, Ni$_{1-x}$Co$_x$PS$_3$, and Fe$_{1-x}$Zn$_x$PS$_3$ etc., has showcased the intriguing and remarkable properties of these alloyed states [11,14–21]. Moreover, substituting nonmagnetic chalcogen atoms, as in MP(*XX′*)$_3$, is expected to fine-tune magnetism through changes in the crystal field splitting of the MX$_6$ octahedra and spin-orbit coupling. Although most research to date has concentrated on mixed *MM′*PX$_3$ compounds or end members, the study of magnetic properties associated with chalcogen-substitution in MP(*XX′*)$_3$ mixed compounds has received relatively little attention. Recent research on chalcogen-substituted MP(*XX′*)$_3$ compounds, specifically MnPS$_{3-x}$Se$_x$, and NiPS$_{3-x}$Se$_x$ by Basnet *et al.* [22] and MnPS$_{3(1-x)}$Se$_{3x}$ by Han *et al.* [23], has demonstrated remarkable alterations in the magnetic properties of these mixed alloyed compounds, as evidenced by magnetic susceptibility measurements.

In addition to diverse static magnetic behaviors, the class of MPX$_3$ material exhibits a variety of quasiparticle excitations, such as magnetic, excitonic, and lattice excitations, along with their mutual interactions. Notable phenomena observed in MPX$_3$ compounds encompass both low-energy and high-energy magnetic excitations, strong magnon-phonon coupling, and coupled magnon-phonon modes, generally known as magnon-polarons [24–28]. Given this, a thorough comprehension of magnon dynamics and its coupling to lattices is essential not only for fundamental research but also for implementing low-energy consumption technologies in magnonic devices. Despite this significance, only room-temperature Raman scattering measurements for MnPSe$_{3-x}$S$_x$ [29] have been conducted, with a lack of detailed temperature-dependent Raman studies. Temperature-dependent Raman scattering has been proven to be a powerful and nondestructive technique to investigate not only lattice vibrations but also various quasiparticle exactions inherent to 2D MPX$_3$ magnetic materials [24,30–32].

In this work, we present a comprehensive temperature- and polarization-dependent Raman study to decipher the impact of chalcogen substitution on the magnetic behaviors in MnPSe$_{3-x}$S$_x$ ($x$= 0.0, 0.5, and 1.5) compounds. We observe an additional phase transition within the long-ranged ordered phase, evident through phonon anomalies and supported by magnetic susceptibility data. This subtle transition is likely linked to spin reorientations arising from additional trigonal distortions of the MnX$_6$ octahedra, facilitated by spin-lattice couplings. In MnPSe$_{3-x}$S$_x$ ($x$=1.5) with maximum chemical disorder, we find that the hybridized character of two-magnon excitations and spin-dependent Raman scatterings weaken due to the presence of competing exchange interactions, highlighting the tunability of magnetic properties through chalcogen substitution.

## 2. Experimental details

Single crystals of MnPSe$_{3-x}$S$_x$ ($x$= 0, 0.5, and 1.5) single crystals were grown using the chemical vapor transport method with iodine as the transport agent. Initially, polycrystalline powders were prepared by the solid-state reaction process. The stoichiometry 3N purity of a selenium slug (Se), phosphorus powder (P), sulfur powder (S), and manganese powders (Mn) were added into a quartz ampoule and sealed under high-vacuum conditions. The mixtures were sintered at 400 ºC and 600 ºC, with intermediate grinding to prepare the single phase of the compounds. For crystal growth, 200 mg of iodine was added to the synthesized powders in a quartz ampoule of 350 mm in length, and sealed under high-vacuum conditions. The tube was kept in a horizontal two-zone furnace and maintained at constant temperatures of 700 and 600 ºC for 200 hrs. After completing the growth process, the furnace temperature was cooled to room temperature at a rate of 2 ºC/min. Finally, high-quality layered single crystals were collected from the cold end of the tubes.

Raman scattering measurements were done using the Princeton SpectraPro HRS-750 spectrometer in backscattering geometry. A 532 nm (2.33 eV) laser was used to excite the Raman spectrum. A laser was kept at a low power of ~300 µW to avoid any local heating and damage to the sample. The incident laser beam was focused on the sample using a 50x objective lens, which was also used to collect the scattered light from the sample. The scattered light was detected using a 1200 grating coupled with an electrically cooled BLAZE Charge Coupled Device detector. Temperature-dependent Raman measurements were carried out using a closed-cycle cryostat (Montana Cryostat) by varying the temperature from 3.5 to 330 K with a temperature accuracy of ± 0.1 K. A waiting time of ~10 minutes was kept after each Raman measurement to ensure temperature stability.

A set of linear polarizer/quarter-wave plates was used to perform polarization-dependent Raman measurements. A vertical analyzer/polarizer was installed in front of the spectrometer to keep the signal constant with respect to the grating orientation. Two quarter-wave plates were utilized: one in the path of incident light and the other in the path of scattered light, to achieve right or left circularly polarized light. The quarter-wave plate in the incident light beam path was adjusted to maintain the helicity of the incident beam on the sample as right ($\sigma^+$) circularly polarized and was kept fixed. The quarter-wave plate in the scattered light beam path was rotated to obtain the helicity of the scattered light beam as either left ($\sigma^-$) or right ($\sigma^+$), resulting in co-circular ($\sigma^+\sigma^+$) and cross-circular ($\sigma^+\sigma^-$) polarized configurations.

Temperature-dependent magnetic susceptibility measurements were carried out using a superconducting quantum interference device vibrating sample magnetometer (SQUID VSM, Quantum Design) with an externally applied magnetic field of $\mu_0 H = 0.02$ T

## 3. Results and Discussion

### A. Raman scattering in MnPSe$_{3-x}$S$_x$

#### A.1 Phonon excitations in MnPSe$_3$

Figure 1(a) shows the unpolarized Raman spectrum of bulk MnPSe$_3$ in a spectral range of 30 - 650 cm$^{-1}$, collected at a temperature of $T$=3.5 K. Section S1 provides information about the crystal and magnetic structures, while Section S2 provides further information about the expected phonons at the Brillouin zone center for both MnPSe$_3$ and MnPS$_3$. The phonon peaks were fitted using a sum of Lorentzian profiles to extract phonon parameters such as the frequency and full width at half maximum (FWHM) of the individual modes. The observed modes are labeled as S1-S24 for convenience and listed in Table S1, along with their frequency and symmetry assignments based on previous studies [24,31,33,34], and our circularly polarized Raman measurements; the circularly polarized Raman spectrum collected at 3.5 K is shown in Fig S2.

In the Raman spectrum of MPX$_3$ materials, despite their complex unit cell, phonons are broadly classified into two categories: (i) external phonons, which are mainly associated with the vibrations of magnetic transition metal ion (Mn$^{2+}$) and lie in the low-frequency range of Raman spectrum, and (ii) internal phonons, which involve symmetric stretching of the P-X and P-P bonds within the $(P_2Se_6)^{4-}$ cluster and are typically situated in the high spectral frequency range. The cutoff frequency that divides the high- and low-frequency regimes varies with the specific M or X atoms. Experimentally, for the case of MnPSe$_3$ and MnPS$_3$, the low-frequency spectral range is expected to be below 140 and 200 cm$^{-1}$, respectively. For MnPSe$_3$, we observed six modes: S1 ($E_g$) and S2 ($E_g$) correspond to the vibrations of $Mn^{2+}$ ions, and S3 ($A_g$), S4 ($E_g$), S6 ($E_g$), and S9 ($A_g$) are associated with the vibrations of the non-magnetic $(P_2Se_6)^{4-}$ cluster. These six peaks have been previously reported in previous Raman studies for MnPSe$_3$ in the frequency range of 70-240 cm$^{-1}$ [24,31,33,34]. Additionally, the broad maximum peaking at 131.1 cm$^{-1}$ (marked by the blue shading in the inset of Fig. 1(a)) is attributed to two-magnon (2M) excitations, being in line with earlier Raman studies [24].

Circularly polarized Raman measurements reveal the $E_g$ character of 2M excitation through its appearance in a cross-circularly polarized configuration. Upon warming, it softens and broadens significantly, first hybridizing with S2 mode around 50-55 K and then with S1 around $T_N$, which will be discussed in more detail later. At low temperatures, three additional weak Raman modes S5 at ~ 162.2 cm$^{-1}$, S7 at ~186.7 cm$^{-1}$, and S8 at ~195.6 cm$^{-1}$ vanish near or above $T_N$, as shown in the inset in Fig. 1(a) and Fig. S5 (a). We assign these $A_g$ modes to magnetically driven activated phonons. It's noteworthy that MnPS$_3$ and MnPSe$_3$ have identical magnetic and crystallographic units, ruling out the possibility of zone-folded phonons due to no folding of the Brillouin zone [35]. Furthermore, no structural changes have been reported for MnPSe$_3$ [8], suggesting that the appearance of these modes could not be attributed to Brillouin zone folding or structural changes. Since these phonons appear below $T_N$, magnetoelastic couplings may be responsible for the S5, S7, and S8 peaks.

The right panel of Fig. 1(a) shows the Raman spectrum of MnPSe$_3$ in a spectral range of 240-650 cm$^{-1}$. The peak S10, assigned to a $A_g$ symmetry, appears narrower than in previous studies, which reported a relatively broader peak at this frequency [36]. We further note that the peak S10 is more intense compared to other modes observed in the 240-650 cm$^{-1}$ range. A group of phonons, S11-S20, is in line with the previous study [36], which become unidentifiable or weaken above $T_N$, see Fig. S4. The observed modes S22 ($E_g$) and S24 ($A_g$) are consistent with those identified in both theoretical and experimental studies [24,37]. S24 may be related to the P-P stretching vibrational mode because this mode was reported at around 500 cm$^{-1}$ in Se-based compounds, while the same mode has been observed at around 600 cm$^{-1}$ in S-based compounds. This indicates that the frequency of this mode is independent of the *M* atoms, yet strongly dependent on the $(P_2X_6)^{4-}$ [35,37]. A broad and weak S21 ($A_g$) mode around 440 cm$^{-1}$ appears to be an overtone of the intense S9 mode, consistent with the selection rule of second-order Raman scattering. Generically, second-order phonons are broader and weaker than the corresponding first-order phonons. The intensity of the n$^{th}$-order modes is generally governed by $g^n$, except for resonant scatterings. Here, *n* is an order of phonon scatterings and *g* is the electron-phonon coupling parameter which is less than one. Therefore, the intensity of higher-order modes is expected to decrease as *n* increases. From Fig. 1(a), it is evident that the S21 is nearly doubled in the frequency of S9, significantly weaker and broader.

**A.2 Phonon excitations in MnPSe$_{2.5}$S$_{0.5}$ and MnPSe$_{1.5}$S$_{1.5}$**

Figures 1(b) and (c) present the unpolarized Raman spectra of MnPSe$_{2.5}$S$_{0.5}$ and MnPSe$_{1.5}$S$_{1.5}$ in a spectral range of 30-650 cm$^{-1}$, collected at 3.5 K, respectively. Surprisingly, distinct changes are observed in these mixed compounds compared to the parent MnPSe$_3$. The most noticeable differences in phonon modes between MnPSe$_3$ and the mixed MnPSe$_{2.5}$S$_{0.5}$ and MnPSe$_{1.5}$S$_{1.5}$ compounds are (i) the linewidth broadening and (ii) the emergence or splitting of certain phonon modes. Specifically, for MnPSe$_{2.5}$S$_{0.5}$ (MnPSe$_{1.5}$S$_{1.5}$) the S1 and S2 modes broaden by ~ 4 (7.5) and ~3 (3.5) cm$^{-1}$, respectively, compared to those of MnPSe$_3$. Similar to S1 and S2, a significant broadening is observed for the S3, S4, and S6 modes in the mixed compounds compared to those of MnPSe$_3$. Interestingly, the S6 mode not only broadens but also splits into two distinct phonon modes, namely S6a and S6b, demonstrating orthogonal behavior under circularly polarized Raman measurements. More specifically, the S6a mode is present in $\sigma^+\sigma^-$ polarization, while the S6b mode is seen in $\sigma^+\sigma^+$ configurations, indicating $E_g$ and $A_g$ symmetries, respectively, see Figs. 1(b and c) and Figs. S2(b and c).

The mode S24 in MnPSe$_3$ splits into three modes: a ~ 505.9, b~ 509.8, and c~ 515.3 cm$^{-1}$ for MnPSe$_{2.5}$S$_{0.5}$ and a ~ 506.6, b~ 510.9, and c~ 516.5 cm$^{-1}$ for MnPSe$_{1.5}$S$_{1.5}$. For MnPSe$_{2.5}$S$_{0.5}$, this splitting persists up to 210 K, while for MnPSe$_{1.5}$S$_{1.5}$, the splitting of S24 modes is observed up to only ~150 K, see Figs. 2 (c) and (f). Surprisingly, the splitting energy differences between the a and b modes and between the b and c modes are identical, $\Delta\omega_{S24(b-a)}$ ~ 4 cm$^{-1}$ and $\Delta\omega_{S24(c-b)}$ ~ 5.5 cm$^{-1}$ for both MnPSe$_{2.5}$S$_{0.5}$ and MnPSe$_{1.5}$S$_{1.5}$. The average frequency is found to be nearly identical: $\omega_{avg}$ ~ 510.3 cm$^{-1}$ for MnPSe$_{2.5}$S$_{0.5}$ and $\omega_{avg}$ ~ 511.3 cm$^{-1}$ for MnPSe$_{1.5}$S$_{1.5}$ but ~ 4-5 cm$^{-1}$ larger than that of the S24 mode in MnPSe$_3$. Notably, S24a is more intense than S24b for MnPSe$_{2.5}$S$_{0.5}$ while S24b is more intense than S24a for MnPSe$_{1.5}$S$_{1.5}$. Furthermore, all split components of S24 are observed in $\sigma^+\sigma^+$ configuration, suggesting $A_g$ symmetry.

As mentioned before, the splitting of the S24 peak, associated with the P-P stretching vibrational mode, is questionable, given its persistence into the phase of short-ranged spin-spin correlations or even beyond the short-ranged phase. Considering the splitting is observed exclusively in the mixed compounds, it may be related to several factors: (i) magnetic/stacking disorder-induced symmetry breaking because the end compounds show different stacking orders and magnetic structures, and (ii) Brillouin zone folding. Since no Brillouin zone folding and thus no zone-folded phonons are expected in the parent MnPSe$_3$ and MnPS$_3$ compounds, the observed phenomenon in the mixed compounds suggests that Brillouin zone folding, due to the differing crystal symmetries of the end members, might activate zone-boundary phonons. Thus, the pronounced splitting of the S24 mode in the mixed compounds is probably linked to either the zone-folding phenomena or magnetic/stacking disorder-induced symmetry breaking.

Additionally, we observed new phonon modes in the mixed MnPSe$_{2.5}$S$_{0.5}$ and MnPSe$_{1.5}$S$_{1.5}$ compounds, labeled P1, P2,…, P12, and N1 and listed in Table S1. These new phonon modes could be related to those of the parent MnPS$_3$ or arise due to local structural distortions in these mixed compounds. Furthermore, we also observe several very weak shoulder modes near intense peaks, marked by an asterisk in Fig. 1(c). The modes P1 at ~ 232 cm$^{-1}$, P2 at ~257.6 cm$^{-1}$, and P3 at ~ 274.9 cm$^{-1}$ closely match the characteristic Raman features of the bulk MnPS$_3$ with $B_g/A_g$, $A_g$ and $B_g/A_g$ symmetry, respectively, attributed to the vibration of the $(P_2S_6)^{4-}$ cluster [38,39]. We note that the selection rule allows $A_g$ mode in co-circular and $B_g$ mode in cross-circular polarized configurations, see Section S3 in Supplementary for more details. We further note that the pairs of quasi-degenerate $A_g$ and $B_g$ modes possess a symmetry similar to $E_g$, as found in phonons around 220-230 cm$^{-1}$ in MnPS$_3$ [38]. In our case, P1 is observed in cross-circular polarization, akin to the $E_g$ mode, indicating that P1 is indeed related to the $B_g/A_g$ mode of MnPS$_3$. On the other hand, P3, observed in co-circular configuration, is in stark contrast with a peak around 275 cm$^{-1}$ as is reported in cross-circular configurations for MnPS$_3$ [38]. MnPSe$_{1.5}$S$_{1.5}$ shows a new broad mode marked by N1 entered at 346.1 cm$^{-1}$, which is absent for MnPSe$_{2.5}$S$_{0.5}$.

In the spectral range of 310-390 cm$^{-1}$, the P7 and P8 modes are grouped into four and three peaks, respectively. For MnPSe$_{2.5}$S$_{0.5}$, the lower side modes of P7a, P7b, and P7c are weaker by an order of magnitude than the P7d mode. On the other hand, the P8b and P8c modes are almost 5 times less intense than the P8a mode. Noticeably, these two P7 and P8 groups become nearly identically intense for the case of MnPSe$_{1.5}$S$_{1.5}$, see Fig. 1(b and c). Furthermore, for MnPSe$_{2.5}$S$_{0.5}$, P7a-c and P8b-c are discernible below $T_N$, see Fig. 2(b). For MnPSe$_{1.5}$S$_{1.5}$, on the other hand, most of the grouped P7 and P8 modes (with a few exceptions) are observed up to our highest recorded temperature, see Fig. 2(e) and Fig. S9.

In addition to the P7 and P8 grouped phonons, we further observe P9, a group of three modes in the spectral 395-425 cm$^{-1}$ in both MnPSe$_{2.5}$S$_{0.5}$ and MnPSe$_{1.5}$S$_{1.5}$ compounds. These P9 phonons are more intense in MnPSe$_{1.5}$S$_{1.5}$ than in MnPSe$_{2.5}$S$_{0.5}$. With the temperature rises, for MnPSe$_{2.5}$S$_{0.5}$, the P9 peaks merge into a broad single peak around 70-80 K and disappear above 150-160 K. For the case of MnPSe$_{1.5}$S$_{1.5}$, the split components of P9 are pronounced at low temperatures, and a single peak is observed up to our highest recorded temperature, see Figs. 2 (c and f) and Fig. S9. Finally, we observe two (P10 and P11) and three (P10-P12) peaks in the spectral range of 525-600 cm$^{-1}$ for MnPSe$_{2.5}$S$_{0.5}$ and MnPSe$_{1.5}$S$_{1.5}$, respectively, corresponding to the high-frequency phonon modes of MnPS$_3$ [37–39].

The appearance of extra phonon peaks, typical for alloyed MPX$_3$ compounds [21,40], may be attributed to multiple factors: (i) spin-spin and spin-phonon coupling, (ii) Brillouin zone folding, (iii) changes in local structural distortions or symmetry breaking, and (iv) quantum interference between discrete phononic state and the electronic excitations due to spin splitting in the band structure [21,40–43]. To simplify the presentation and avoid confusion, from now on, we will use our simple phonon mode notation, unless symmetry of phonon mode is required, see Table S1 for the symmetry assignments of the modes.

**B. Two-Magnon Raman scattering**

Now we turn to 2M excitations and their temperature and composition dependence. In antiferromagnets, 2M Raman scattering occurs through the excitations of two magnons with equal but opposite momentum [44,45]. Since the 2M scattering processes are dictated by the exchange scattering mechanism [45–47], its scattering intensity can be much stronger than the one-magnon one, bearing information about spin dynamics and magnetic exchange interactions.

Figures 3 (a-c) show the unpolarized 2D color contour maps of the Raman intensity versus Raman shift across a temperature range of 3.5-330 K for (a) $MnPSe_3$, (b) $MnPSe_{2.5}S_{0.5}$, and (c) $MnPSe_{1.5}S_{1.5}$ in a frequency range of 30-175 cm$^{-1}$. The 2M excitation is marked by a red arrow. In the unpolarized Raman spectrum of $MnPSe_3$, the 2M feature is well-resolved and separated from phonon modes at the lowest recorded temperature, while it appears as a sideband in $MnPSe_{2.5}S_{0.5}$. On the other hand, the 2M overlaps with phonons for $MnPSe_{1.5}S_{1.5}$, see Figs.1(c) and 3(c). Notably, the existence of the 2M excitations for $MnPSe_{1.5}S_{1.5}$ can be deduced from the thermal evolution of the circularly polarized Raman spectrum as shown in Fig. 3(f) and Fig. S6. With increasing temperature toward $T_N$, the 2M excitations in all three compounds soften and broaden rapidly. This behavior contrasts with conventional 2M excitations, which persist up to several $T_N$ as a form of paramagnons. The strong damping and renormalization are attributed to hybridization with the S2 mode, as evinced by the asymmetric line shape of the S2 mode. We note that the hybridization of 2M with S1 is not directly visible, see Fig. 3 and Figs. S5 and S6.

To gain a quantitative insight into the hybridization of the 2M excitations with the S2 and S1 modes, we plot the temperature dependence of their frequency and FWHM in Fig. 4. The hybridization between two quasiparticles manifests through shifts in frequency and linewidth broadening as the phonons' energy and lifetime are renormalized and shortened due to the additional relaxation channels provided by their interaction with 2M excitations.

In the case of MnPSe$_3$, the S2 mode shows a softening in frequency below 50 K, accompanying a maximum in FWHM, as indicated by a solid red vertical line in Fig. 4 (a). Similarly, the S1 mode experiences energy renormalization and line broadening around $T_N$, see a red vertical dashed line in Fig. 4 (a). Above $T_N$, the FWHM of the S1 mode decreases, reaching its minimum at around 100 K. This phenomenon can be rationalized by the interaction between the damped paramagnons in the paramagnetic state and the lower-frequency S1 mode, resulting in a wider temperature range of hybridization [47]. Noteworthy is that the enhanced FWHM in the vicinity of the ordered phase is described by the Lorentzian line shape, see a solid red line in Fig. 4 (d), with peak centers at 52.8 K and ~79.5 K for S2 and S1, respectively. Similar anomalies in the frequencies and FWHMs of the S2 and S1 are observed in MnPSe$_{2.5}$S$_{0.5}$, confirming the existence of the same hybridization. In MnPSe$_{1.5}$S$_{1.5}$, while the hybridization characteristic remains discernible for the S2 mode, it becomes less apparent for the S1 mode due to chemical disorder, which already shortens its lifetime.

Next, we attempt to deduce magnetic exchange interactions from the energies of the 2M peaks based on the expression $\omega_{2m} = J(2zS-1)$, where $\omega_{2m}$, $S = 5/2$, and $z = 3$ denote the peak frequency of 2M, spin number, and the number of nearest-neighbor spins, respectively. These materials comprise three intraplanar exchange coupling parameters: $J1$, $J2$, and $J3$, corresponding to the first, second, and third nearest-neighbor exchange interactions, respectively. While $J1$ and $J3$ possess antiferromagnetic exchange interactions, $J2$ has ferromagnetic interactions, see Fig. S1. Since the single energy of 2M is not sufficient to determine all the magnetic exchange parameters, the determined $J$ should be regarded as a sum of $J1$, $J2$, and $J3$. From the 2M peak energies observed at 130.6, 136.9, and 140.7 cm$^{-1}$ for MnPSe$_3$, MnPSe$_{2.5}$S$_{0.5}$, and MnPSe$_{1.5}$S$_{1.5}$, respectively, $J$ is estimated to be 1.15, 1.20, and 1.24 meV. Overall, the estimated values of $J$ appear somewhat larger than $J$ ($=J1+J2+J3$)=0.67 meV for MnPSe$_3$ and 0.78 meV for MnPSe$_{1.5}$S$_{1.5}$, derived from inelastic neutron scattering [8,48]. This discrepancy arises from the underestimation of $z=6$ numbers for $J2$ and $J3$. Going from MnPSe$_3$ to MnPS$_3$, an increase in exchange interaction is explained by the increasing chemical pressure due to the smaller size of S compared to the Se atoms.

**C. Emergent multiple phase transitions: Temperature dependence of phonon modes**

In the following, we scrutinize the magnetic phase through spin-phonon coupling. Figure 5 illustrates the temperature dependence of frequency and FWHM for several prominent modes: panel (a) S9 and P2-P6 for MnPSe$_{2.5}$S$_{0.5}$ and panel (b) S9, P1-P3, P5, and P6 for MnPSe$_{1.5}$S$_{1.5}$. The low-frequency S1 and S2 modes are shown in Fig. 4 for all three compounds. Figure S7 shows the temperature-dependent frequency and FWHM of S3, S4, S6, and S9 in MnPSe$_3$. Additionally, Figs. S8 (a) and (b) display the $T$-dependent frequency and FWHM for S3, S4, and S6 of MnPSe$_{2.5}$S$_{0.5}$ and MnPSe$_{1.5}$S$_{1.5}$, respectively.

From Figs. 4, and 5, and Figs. S7, and S8, for all three compounds, we observed that a phonon softening and broadening above $T_N$ or short-range ordering can be captured within a three-phonon anharmonicity model [49]: $\omega(T) = \omega_0 + A(1+\frac{2}{e^x-1})$ and $\Gamma(T) = \Gamma_0 + C(1+\frac{2}{e^x-1})$, where $x = \hbar\omega/2k_BT$ and $\omega_0$ ($\Gamma_0$) represent the frequency (FWHM) of the phonon mode at zero temperature. The coefficients A and C are the cubic anharmonicity fitting parameters of the three-phonon anharmonicity contributions to $\omega(T)$ and $\Gamma(T)$, respectively. The solid red lines in Figs. 5 (a and b) and Fig. S7 validate the adequacy of our employed model.

Now we inspect the *T* dependence of the phonon modes below $T_N$ or the short-ranged ordered phase. From Fig. 4, and Fig. S7 for MnPSe$_3$, the following observations are noteworthy: (i) the frequency of the S1 (S2) modes softens with decreasing temperature from $T_N$ (~50 K) to 40 K (20 K), followed by a slight increase (nearly constant) upon further cooling, (ii) the frequency of the S3 mode remains nearly constant, while the frequencies of the S4 and S6 modes decrease with lowering temperature, (iii) the frequency of S9 slightly increases with decreasing temperature, and (iv) the FWHM of the S1 and S2 modes show a maximum, as discussed in more detail in Section 3 (B), while the FWHM of the S3, S4, S6, and S9 modes show conventional temperature dependence. The observed anomalies allude to the presence of additional mechanisms.

Within the spin-phonon coupling mechanism, the bare phonon frequency $\omega_0$ is renormalized by spin-spin correlations $\langle S_i.S_j \rangle$ between the i$^{th}$ and j$^{th}$ sites of the magnetic ions through spin-phonon coupling $\lambda$: $\Delta\omega_{sp-ph} = \omega_{sp-ph}(T) - \omega_0 = \lambda * \langle \vec{S_i}.\vec{S_j} \rangle$. In the mean-field approximation, we obtain $\langle S_i.S_j \rangle = -S^2 \phi(T)$, where $\phi(T) = 1 - \left(\frac{T}{T_N}\right)^\gamma$ is the order parameter with the critical exponent $\gamma$. For 2D magnetic materials like the studied MPX$_3$, short-ranged spin-spin correlations survive up to several $T_N$ [14,23]. Indeed, the frequency anomalies are discernible up to 80-100 K, see Fig. S7 for certain phonons. The observed anomalous phonon softening on cooling from 100 to ~20 K results from the spin-phonon contributions [31]. Remarkably, below ~ 20 K, we further observe a renormalization in the phonon self-energy, clearly visible in both the frequency and FWHM of S1 and S2, as well as in the intensity of all phonon modes, as depicted in Fig. 4 and Fig. 6. Similar anomalies in phonon frequencies are seen for MnPSe$_{2.5}$S$_{0.5}$ (S9, and P2-P6) and MnPSe$_{1.5}$S$_{1.5}$ (S9, P1-P3, and P5-P6) as well.

However, the anomaly is not obvious in the FWHM of the phonon for both MnPSe$_{2.5}$S$_{0.5}$ and MnPSe$_{1.5}$S$_{1.5}$. Instead, the FWHMs show continuous narrowing across $T_N$ with a small change in slope, similar to the S3, S4, S6, and S9 modes of MnPSe$_3$, see Fig. 5 and Fig. S7.

For MnPSe$_{2.5}$S$_{0.5}$, phonon frequencies show distinct changes in slope with decreasing temperature through ~ 82-85 K, suggesting a transition into the long-ranged ordered phase. Intriguingly, upon further cooling within this ordered phase, a noticeable increase in the phonon frequency around 33 K indicates a crossover into another phase, as indicated by the green shaded area in Fig. 5 (a). Similarly, for MnPSe$_{1.5}$S$_{1.5}$, the frequency experiences a slope change, while certain modes show a downturn or an upturn at ~117 K, signaling that the system enters into a short-range ordered phase. With further lowering temperature below 117 K, we observe another upturn in the frequency around 80 K, followed by a subsequent downturn around 50 K. The successive transitions suggest the onset of yet another phase inside the long-range ordered state, see the green shaded area in Fig. 5(b). The self-energy renormalization as a function of temperature supports the emergence of a new phase for all three compounds at ~17.5, ~ 33, and ~ 50 K below the long-ranged ordered temperatures for MnPSe$_3$, MnPSe$_{2.5}$S$_{0.5}$, MnPSe$_{1.5}$S$_{1.5}$, respectively.

To further confirm the observation of multiple magnetic phase transitions, we measured magnetic susceptibility as a function of temperature for both $\mu_0 H //ab$ (black line in Fig. S10) and $\mu_0 H //c$ (red line in Fig. S10) under an external field of 0.02 T. We observe a clear peak centered around ~ 37 K for the case of MnPSe$_{1.5}$S$_{1.5}$, which disappears under a higher magnetic field. For other samples, the secondary phase is not evident due to the low-$T$ Curie tail. Similar anomalous peaks have also been reported in previous magnetic susceptibility [23,50], but their origin is the subject of ongoing investigations.

Here, we recall that the trigonal distortion of MnX$_6$ octahedra determines a magnetic anisotropy, which modulates spin dimensionality. Both MnPSe$_3$ and MnPS$_3$ exhibit XY anisotropy [7,13,51]. A 2D XY system undergoes a topological phase transition, known as the Berezinskii-Kosterlitz-Thouless (BKT) transition, at a finite temperature due to the binding of magnetic vortex-antivortex pairs [52]. The BKT mechanism has been asserted in several MPX$_3$ members, even in bulk forms [7,51,53]. Nonetheless, the bulk MnPSe$_{3-x}$S$_x$ is known to stabilize a long-range magnetic order thanks to interlayer interactions. We further find no indication that the 2M excitation evolves into topological excitations at the temperatures where the phonon parameters show anomalies, see Fig. 3. Possibly, the subtle trigonal distortion of MnX$_6$ octahedra might cause spin reorientations by changing the magnetic anisotropy, leading to the renormalization of the phonon self-energy. Future investigations on monolayer MnPSe$_{3-x}$S$_x$ are needed to clarify how the magnetic anomalies evolve to the putative BKT transition in the 2D limit, if present.

### D. Tunable spin-dependent Raman scattering

In this section, we address the $T$-dependent Raman scattering intensity $I(T)$ of phonons in the magnetically ordered phase. According to the Suzuki and Kamimura (SK) theory, the integrated $T$-dependent Raman scattering intensity of phonons is given as [54]

$$I(T) = (n+1)\left[\left|R + M\frac{\langle S_i \cdot S_j \rangle}{S^2}\right|^2 + |K|^2 \langle S_z^2 \rangle\right], \quad (1)$$

where the term $(n+1)$ represents the Bose-Einstein factor. The terms associated with the coefficients $R$ and $M$ account for the spin-independent and spin-dependent contribution to $I(T)$, respectively. The last term $K^2 \langle S_z^2 \rangle$, associated with the spins of single ions, could be neglected because its magnitude is two orders of magnitude smaller than that of $R$ and $M$. Assuming the spin-spin correlations $\langle S_i.S_j \rangle = -S^2\phi(T)$, the SK formula is simplified to

$$I(T) = \left| R - M \left[ 1 - \left( \frac{T}{T_N} \right)^\gamma \right] \right|^2. \qquad (2)$$

Figures 6 (a-c) present the $I(T)$ of selected phonons for MnPSe$_{3-x}$S$_x$ ($x$ = 0, 0.5, and 1.5). For MnPSe$_3$ and MnPSe$_{2.5}$S$_{0.5}$, an abrupt decrease in intensity is observed with increasing temperature up to $T_N$, followed by a slight decrease up to a certain temperature above $T_N$. With a further increase in temperature, the intensity remains nearly constant. For MnPSe$_{1.5}$S$_{1.5}$, on the other hand, no abrupt change in intensity with temperature takes place. Instead, it shows an exponential decrease with increasing temperature in contrast to MnPSe$_3$ and MnPSe$_{2.5}$S$_{0.5}$. For MnPSe$_3$, the S1 and S2 modes involving the $M^{2+}$ ions are subject to the modulation of the phonon intensity due to the spin ordering. However, other modes such as S3, S4, S6, S9, etc., are associated with the non-magnetic $(P_2Se_6)^{4-}$ clusters and also exhibit similar $T$-dependent intensity as S1 and S2. The analogous $T$-dependent intensities of non-magnetic and magnetic ion vibrational modes can be understood by considering that the spin-dependent Raman scattering arises from $d$-electron transfer between magnetic ions through super-exchange paths involving non-magnetic ions.

Figure 6 (d) sketches a schematic representation of the spin-dependent Raman scattering process. Through this electron-phonon interaction, the spin ordering also has a significant impact on the vibrational modes associated with non-magnetic ions. In MnPSe$_3$, the non-magnetic $(P_2Se_6)^{4-}$ clusters participate in Kramers-Anderson super-exchange pathways between neighboring magnetic ions [31,55,56], which explains the $T$-dependent phonon intensity of S3, S4, S6 S9. We note that these modes involve the non-magnetic $(P_2Se_6)^{4-}$ and display similar behavior to S1 and S2. MnPSe$_{2.5}$S$_{0.5}$ exhibits phonon intensity behavior, akin to MnPSe$_3$. However, the *T*-dependent intensity trend for MnPSe$_{1.5}$S$_{1.5}$ differs markedly, displaying a nearly exponential behavior rather than the abrupt change observed in MnPSe$_3$ and MnPSe$_{2.5}$S$_{0.5}$. This suggests a weakening of spin-dependent Raman scattering in MnPSe$_{1.5}$S$_{1.5}$.

For quantitative analysis, we fitted *I*(*T*) of the intense phonon modes to Eq. (2) in the magnetically ordered phase, see Fig. S11 and Table S2 for the fitting parameters. The spin-independent *R* term increases, while the spin-independent *M* term decreases with increasing S concentration, indicating a weakening in the spin-dependent Raman scattering. Variations in spin-dependent Raman scattering have also been reported as a function of temperature and composition [17]. The weakening or diminishing of the spin-dependent Raman scattering in MnPSe$_{1.5}$S$_{1.5}$ may result from enhanced chemical disorder and the presence of competing exchange interactions through $(P_2(Se_{1-x}S_x)_6)^{4-}$ clusters. We note that the parent compounds MnPSe$_3$ and MnPS$_3$ exhibit the in-plane and out-of-plane spin alignments, respectively. Consequently, in MnPSe$_{1.5}$S$_{1.5}$ two super-exchange processes become frustrated, destructively interfering between two competing spin-dependent scattering processes.

## 4. Conclusion

In summary, we have reported a detailed Raman scattering study of MnPSe$_{3-x}$S$_x$ (*x* = 0, 0.5, and 1.5) to understand the intricate confluence of lattice and magnetic excitations as functions of temperature and chemical composition. Probing two-magnon excitations across all investigated compounds reveals that the strength of their hybridization with S2 to S1 phonon modes is notably affected by chalcogen substitution. Through Raman spectroscopy and magnetic susceptibility measurements, we identify a sequence of three-phase transitions with decreasing temperature: short- and long-range magnetically ordered phases, and a distinct low-temperature phase inside the long-range ordered state, associated with spin reorientations or magnetic instabilities inherent to two-dimensional topological state. We further find the significant role of spin dynamics and the Kramers-Anderson super-exchange pathways in modulating spin-dependent phonon scattering intensities. Our work demonstrates the potential of chalcogen substitution as a tool for tailoring their magnetic and lattice characteristics as well as their mutual coupling through engineering $(P_2(Se_{1-x}S_x)_6)^{4-}$ clusters.

**Acknowledgments:**

This work was supported by the National Research Foundation (NRF) of Korea (Grant Nos. 2020R1A5A1016518, 2022R1A2C1003959, and RS-2023-00209121). R.S. acknowledges the financial support provided by the Ministry of Science and Technology in Taiwan under project numbers NSTC 111-2124-M-001-007, NSTC-110-2112-M-001-065-MY3, and NSTC111-2124-M-A49-009 and Academia Sinica for the budget of AS-iMATE-111-12.

**References:**

[1] N. D. Mermin and H. Wagner, Absence of Ferromagnetism or Antiferromagnetism in One- or Two-Dimensional Isotropic Heisenberg Models, Phys. Rev. Lett. **17**, 1133 (1966).

[2] B. Huang, G. Clark, E. N. Moratalla, D. R. Klein, R. Cheng, K. L. Seyler, D. Zhong, E. Schmidgall, M. A. McGuire, D. H. Cobden, W. Yao, D. Xiao, P. J. Herrero, and X. Xu, Layer-Dependent Ferromagnetism in a van Der Waals Crystal down to the Monolayer Limit, Nature **546**, 270 (2017).

[3] C. Gong, L. Li, Z. Li, H. Ji, A. Stern, Y. Xia, T. Cao, W. Bao, C. Wang, Y. Wang, Z. Q. Qiu, R. J. Cava, S. G. Louie, J. Xia, and Xiang Zhang, Discovery of Intrinsic Ferromagnetism in Two-Dimensional van Der Waals Crystals, Nature **546**, 265 (2017).

[4] X. Jiang, Q. Liu, J. Xing, N. Liu, Y. Guo, Z. Liu, and J. Zhao, Recent Progress on 2D Magnets: Fundamental Mechanism, Structural Design and Modification, Appl. Phys. Rev. **8**, 031305 (2021).

[5] Q. H. Wang, A. B. Pinto, M. Blei, A. H. Dismukes, A. Hamo, S. Jenkins, M. Koperski, Y. Liu, Q. C. Sun, E. J. Telford et al., The Magnetic Genome of Two-Dimensional van Der Waals Materials, ACS Nano **16**, 6960(2021).

[6] J. U. Lee, S. Lee, J. H. Ryoo, S. Kang, T. Y. Kim, P. Kim, C. H. Park, J. G. Park, and H. Cheong, Ising-Type Magnetic Ordering in Atomically Thin $FePS_3$, Nano Lett. **16**, 7433 (2016).

[7] S. Chaudhuri, C. N. Kuo, Y. S. Chen, C. S. Lue, and J. G. Lin, Low-Temperature Magnetic Order Rearrangement in the Layered van Der Waals Compound $MnPS_3$, Phys. Rev. B **106**, 094416 (2022).

[8] S. Calder, A. V. Haglund, A. I. Kolesnikov, and D. Mandrus, Magnetic Exchange Interactions in the van Der Waals Layered Antiferromagnet $MnPSe_3$, Phys. Rev. B **103**, 024414 (2021).

[9] K. Kim, S. Y. Lim, J. U. Lee, S. Lee, T. Y. Kim, K. Park, G. S. Jeon, C. H. Park, J. G. Park, and H. Cheong, Suppression of Magnetic Ordering in XXZ-Type Antiferromagnetic Monolayer $NiPS_3$, Nat. Commun. **10**, 345 (2019).

[10] P. A. Joy and S. Vasudevan, Magnetism in the Layered Transition-Metal Thiophosphates $MPS_3$ (M=Mn, Fe, and Ni), Phys. Rev. B **46**, 5425 (1992).

[11] S. Selter, Y. Shemerliuk, M. I. Sturza, A. U. B. Wolter, B. Büchner, and S. Aswartham, Crystal Growth and Anisotropic Magnetic Properties of Quasi-Two-Dimensional $(Fe_{1-x}Ni_x)_2P_2S_6$, Phys. Rev. Mater. **5**, 073401 (2021).

[12] D. Lançon, H. C. Walker, E. Ressouche, B. Ouladdiaf, K. C. Rule, G. J. McIntyre, T. J. Hicks, H. M. Rønnow, and A. R. Wildes, Magnetic Structure and Magnon Dynamics of the Quasi-Two-Dimensional Antiferromagnet $FePS_3$, Phys. Rev. B **94**, 214407 (2016).

[13] P. Jeevanandam and S. Vasudevan, Magnetism in $MnPSe_3$: A Layered 3d5 Antiferromagnet with Unusually Large XY Anisotropy, J. Phys. Condens. Matter **11**, 3563 (1999).

[14] A. Bhutani, J. L. Zuo, R. D. McAuliffe, C. R. Dela Cruz, and D. P. Shoemaker, Strong Anisotropy in the Mixed Antiferromagnetic System $Mn_{1-x}Fe_xPSe_3$, Phys. Rev. Mater. **4**, 34411 (2020).

[15] J. N. Graham, M. J. Coak, S. Son, E. Suard, J. G. Park, L. Clark, and A. R. Wildes, Local Nuclear and Magnetic Order in the Two-Dimensional Spin Glass $Mn_{0.5}Fe_{0.5}PS_3$, Phys. Rev. Mater. **4**, 084401 (2020).

[16] R. Basnet, A. Wegner, K. Pandey, S. Storment, and J. Hu, Highly Sensitive Spin-Flop Transition in Antiferromagnetic van Der Waals Material $MPS_3$ (M=Ni and Mn), Phys. Rev. Mater. **5**, 064413 (2021).

[17] S. Lee, J. Park, Y. Choi, K. Raju, W. T. Chen, R. Sankar, and K. Y. Choi, Chemical Tuning of Magnetic Anisotropy and Correlations in $Ni_{1-x}Fe_xPS_3$, Phys. Rev. B **104**, 174412 (2021).

[18] N. Chandrasekharan and S. Vasudevan, Dilution of a Layered Antiferromagnet: Magnetism, Phys. Rev. B **54**, 14903 (1996).

[19] V. Manrı, P. Barahona, and O. Pen, Physical Properties of the Cation-Mixed M J MPS 3 Phases, **35**, 1889 (2000).

[20] F. Wang, N. Mathur, A. N. Janes, H. Sheng, P. He, X. Zheng, P. Yu, A. J. DeRuiter, J. R. Schmidt, J. He, and S. Jin, Defect-Mediated Ferromagnetism in Correlated Two-Dimensional Transition Metal Phosphorus Trisulfides, Sci. Adv. **7**, 1 (2021).

[21] N. Khan, D. Kumar, V. Kumar, Y. Shemerliuk, S. Selter, B. Büchner, K. Pal, S. Aswartham, and P. Kumar, Interplay of topology and antiferromagnetic order in two-dimensional van der Waals crystals of $(Ni_xFe_{1-x})_2P_2S_6$, https://arxiv.org/abs/2312.01098.

[22] R. Basnet, K. M. Kotur, M. Rybak, C. Stephenson, S. Bishop, C. Autieri, M. Birowska, and J. Hu, Controlling Magnetic Exchange and Anisotropy by Nonmagnetic Ligand Substitution in Layered $MPX_3$ (M=Ni, Mn; X= S, Se), Phys. Rev. Res. **4**, 023256 (2022).

[23] H. Han, H. Lin, W. Gan, Y. Liu, R. Xiao, L. Zhang, Y. Li, C. Zhang, and H. Li, Emergent Mixed Antiferromagnetic State in $MnPS_{3(1-x)}Se_{3x}$, Appl. Phys. Lett. **122**, 033101 (2023).

[24] T. T. Mai, K. F. Garrity, A. McCreary, J. Argo, J. R. Simpson, V. Doan-Nguyen, R. V. Aguilar, and A. R. Hight Walker, Magnon-Phonon Hybridization in 2D Antiferromagnet $MnPSe_3$, Sci. Adv. **7**, 1 (2021).

[25] S. Liu, A. G. del Águila, D. Bhowmick, C. K. Gan, T. Thu Ha Do, M. A. Prosnikov, D. Sedmidubský, Z. Sofer, P. C. M. Christianen, P. Sengupta, and Q. Xiong, Direct Observation of Magnon-Phonon Strong Coupling in Two-Dimensional Antiferromagnet at High Magnetic Fields, Phys. Rev. Lett. **127**, 097401 (2021).

[26] D. Jana, P. Kapuscinski, I. Mohelsky, D. Vaclavkova, I. Breslavetz, M. Orlita, C. Faugeras, and M. Potemski, Magnon Gap Excitations and Spin-Entangled Optical Transition in the van Der Waals Antiferromagnet $NiPS_3$, Phys. Rev. B **108**, 115149(2023).

[27] J. Luo, S. Li, Z. Ye, R. Xu, H. Yan, J. Zhang, G. Ye, L. Chen, D. Hu, X. Teng, W. A. Smith, B. I. Yakobson, P. Dai, A. H. Nevidomskyy, R.He, and H. Zhu, Evidence for Topological Magnon-Phonon Hybridization in a 2D Antiferromagnet down to the Monolayer Limit, Nano Lett. **23**, 2023 (2023).

[28] A. McCreary, J. R. Simpson, T. T. Mai, R. D. McMichael, J. E. Douglas, N. Butch, C. Dennis, R. Valdés Aguilar, and A. R. Hight Walker, Quasi-Two-Dimensional Magnon Identification in Antiferromagnetic $FePS_3$ via Magneto-Raman Spectroscopy, Phys. Rev. B **101**, 64416 (2020).

[29] X. Hou, X. Zhang, Q. Ma, X. Tang, Q. Hao, Y. Cheng, and T. Qiu, Alloy Engineering in Few-Layer Manganese Phosphorus Trichalcogenides for Surface-Enhanced Raman Scattering, Adv. Funct. Mater. **30**, 191171 (2020).

[30] V. Kumar, D. Kumar, B. Singh, Y. Shemerliuk, M. Behnami, B. Büchner, S. Aswartham, and P. Kumar, Fluctuating Fractionalized Spins in Quasi-Two-Dimensional Magnetic $V_{0.85}PS_3$, Phys. Rev. B **107**, 094417 (2023).

[31] D. J. Gillard, D. Wolverson, O. M. Hutchings, and A. I. Tartakovskii, Spin-Order-Dependent Magneto-Elastic Coupling in Two Dimensional Antiferromagnetic $MnPSe_3$ Observed through Raman Spectroscopy, Npj 2D Mater. Appl. **8**, 1 (2024).

[32] Y. J. Sun, Q. H. Tan, X. L. Liu, Y. F. Gao, and J. Zhang, Probing the Magnetic Ordering of Antiferromagnetic $MnPS_3$ by Raman Spectroscopy, J. Phys. Chem. Lett. **10**, 3087 (2019).

[33] M. Chisa, S. Tomoyuki, T. Yoshiko, and K. Koh, Raman scattering in the two-dimensional antiferromagnet $MnPSe_3$, J. Phys Condens. Matter **5**, 623 (1993).

[34] P. Liu, Z. Xu, H. Huang, J. Li, C. Feng, M. Huang, M. Zhu, Z. Wang, Z. Zhang, D. Hou, Y. Lu, and B. Xiang, Exploring the Magnetic Ordering in Atomically Thin Antiferromagnetic $MnPSe_3$ by Raman Spectroscopy, J. Alloys Compd. **828**, 154432 (2020).

[35] M. Bernasconi, G. L. Marra, G. Benedek, L. Miglio, M. Jouanne, C. Julien, M. Scagliotti, and M. Balkanski, Lattice Dynamics of Layered $MPX_3$ (M=Mn,Fe,Ni,Zn; X=S,Se) Compounds, Phys. Rev. B **38**, 12089 (1988).

[36] J. Qiu, W. J. Huang, Y. J. Lee, S. Kim, X. B. Chen, and I. S. Yang, Raman Study of the Spin–Lattice Excitations of the Layered Antiferromagnets $MPSe_3$ (M = Fe, Mn), Results Phys. **56**, 107256 (2024).

[37] A. Hashemi, H. P. Komsa, M. Puska, and A. V. Krasheninnikov, Vibrational Properties of Metal Phosphorus Trichalcogenides from First-Principles Calculations, J. Phys. Chem. C **121**, 27207 (2017).

[38] K. Kim, S. Y. Lim, J. Kim, J.U. Lee, S. Lee, P. Kim, K. Park, S. Son, C. H Park, J. G. Park, Antiferromagnetic Ordering in van Der Waals 2D Magnetic Material $MnPS_3$ Probed by Raman Spectroscopy, 2D Mater. **6**, 041001 (2019).

[39] D. Vaclavkova, A. Delhomme, C. Faugeras, M. Potemski, A. Bogucki, J. Suffczyński, P. Kossacki, A. R. Wildes, B. Grémaud, and A. Saúl, Magnetoelastic Interaction in the Two-Dimensional Magnetic Material $MnPS_3$ studied by First Principles Calculations and Raman Experiments, 2D Mater. **7**, 035030(2020).

[40] R. Oliva, E. Ritov, F. Horani, I. Etxebarria, A. K. Budniak, Y. Amouyal, E. Lifshitz, and M. Guennou, Lattice Dynamics and In-Plane Antiferromagnetism in $Mn_xZn_{1-x}PS_3$ across the Entire Composition Range, Phys. Rev. B **107**, 104415 (2023).

[41] M. Balkanski, M. Jouanne, G. Ouvrard, and M. Scagliotti, Effects Due to Spin Ordering in Layered $MPX_3$ Compounds Revealed by Inelastic Light Scattering, J. Phys. C Solid State Phys. **20**, 4397 (1987).

[42] G. Güntherodt, Light scattering in magnetic semiconductors, J. Magn. Magn. Mater. **11**, 394 (1979).

[43] M. Balkanski, K. P. Jain, R. Beserman, and M. Jouanne, Theory of Interference Distortion of Raman Scattering Line Shapes in Semiconductors, Phys. Rev. B **12**, 4328 (1975).

[44] R. Loudon, The Raman Effect in Crystals, Adv. Phys. **50**, 813 (2001).

[45] P. A. Fleury and R. Loudon, Zs, **166**, 514(1968).

[46] A. Menth, E. Buehler, and T. H. Geballe, Physical Review Letters 17, **22**, 295 (1969).

[47] P. A. Fleury, Paramagnetic Spin Waves and Correlation Functions in NiF$_2$, Phys. Rev. **180**, 591 (1969).

[48] R. Baral, A. V. Haglund, J. Liu, A. I. Kolesnikov, D. Mandrus, and S. Calder, Local Spin Structure in the Layered van Der Waals Materials MnPS$_x$Se$_{3-x}$, https://doi.org/10.48550/arXiv.2404.02328.

[49] P. G. Klemens, Anharmonic Decay of Optical Phonon in Diamond, Phys. Rev. B **11**, 3206 (1975).

[50] X. Yan, X. Chen, and J. Qin, Synthesis and Magnetic Properties of Layered MnPS$_x$Se$_{3-x}$ (0 < x < 3) and Corresponding Intercalation Compounds of 2,2′-Bipyridine, Mater. Res. Bull. **46**, 235 (2011).

[51] A. R. Wildes, H. M. Rønnow, B. Roessli, M. J. Harris, and K. W. Godfrey, Static and Dynamic Critical Properties of the Quasi-Two-Dimensional Antiferromagnet MnPS$_3$, Phys. Rev. B, **74**, 094422 (2006).

[52] J. M. kosterlitz and D. J. Thouless, Ordering, Metastability and Phase Transitions in Two-Dimensional Systems, J. Phys. C Solid State Phys. **6**, 1181 (1973).

[53] G. Liao, S. Zhang, P. Cui, and Z. Zhang, Tunable Meron Pair Excitations and Berezinskii-Kosterlitz-Thouless Phase Transitions in the Monolayer Antiferromagnet MnPS E3, Phys. Rev. B **109**, 100403 (2024).

[54] N. Suzuki, H. Kamimura, Theory of spin-dependent phonon Raman scattering in magnetic crystals, J. Phys. Soc. Jpn. **35**, 985 (1973)

[55] P. W. Anderson, Antiferromagnetism. Theory of Superexchange Interaction, Phys. Rev. **79**, 350 (1950).

[56] H. A. Kramers, L'interaction Entre les Atomes Magnétogènes dans un Cristal Paramagnétique, Physica **1**, 182 (1934).

**Figure**

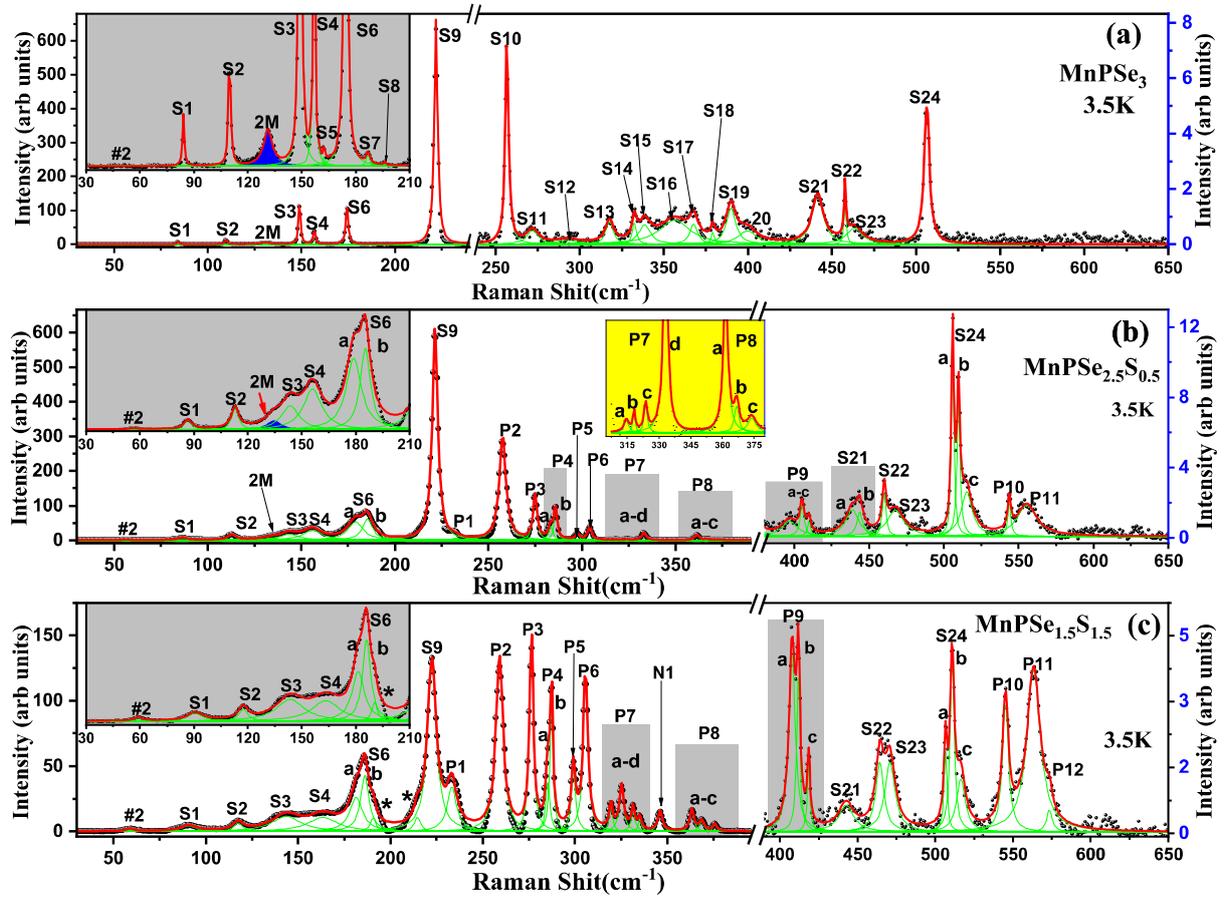

**Figure 1:** Unpolarized Raman spectrum of (a) MnPSe$_3$, (b) MnPSe$_{2.5}$S$_{0.5}$, and (c) MnPSe$_{1.5}$S$_{1.5}$ in the spectral range of 30-650 cm$^{-1}$ collected at 3.5 K. The solid red line shows the total sum of Lorentzian fit, and the thin green lines show the individual fits of the Raman features. The observed phonon modes are labeled as S1-S24, P1-P10, and N1. Insets (a), (b), and (c) in gray are the amplified spectrum for the spectral range of 30-210 cm$^{-1}$. Inset (b) in yellow is the amplified spectrum for the spectral range of 305-380 cm$^{-1}$. The observed peak (blue shaded) is attributed to the two-magnon (2M) excitations. For the case of MnPSe$_{1.5}$S$_{1.5}$, the two-magnon signal is not clearly visible due to very close to intense and broad S3 mode.

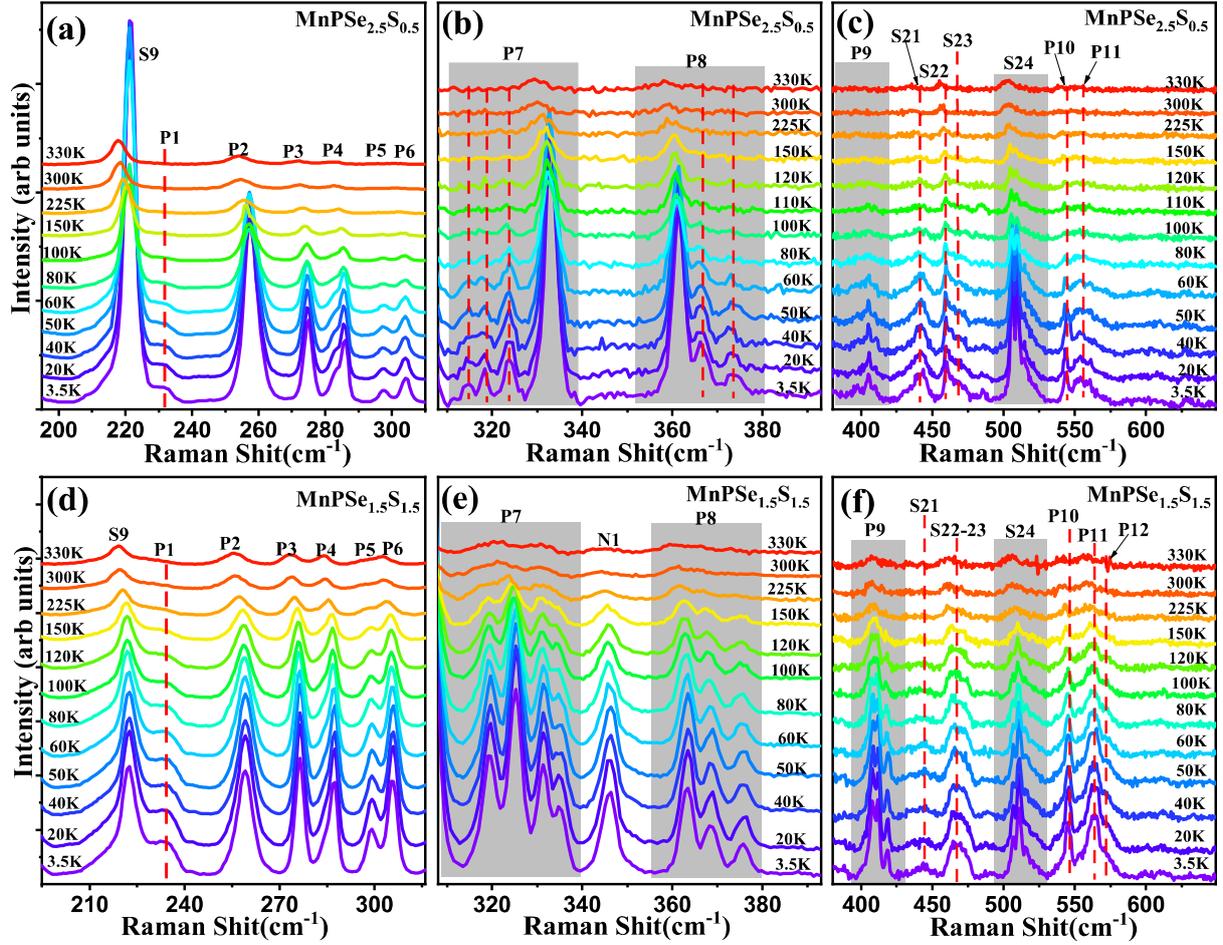

**Figure 2:** (a), (b), and (c) Temperature evolution of the unpolarized Raman spectrum of MnPSe$_{2.5}$S$_{0.5}$ in the frequency range of 190-310 cm$^{-1}$, 310-390 cm$^{-1}$, and 380-650 cm$^{-1}$, respectively. (d), (e), and (f) Temperature evolution of the Raman spectrum of MnPSe$_{1.5}$S$_{1.5}$ in a frequency range of 190-310 cm$^{-1}$, 310-390 cm$^{-1}$, and 380-650 cm$^{-1}$, respectively.

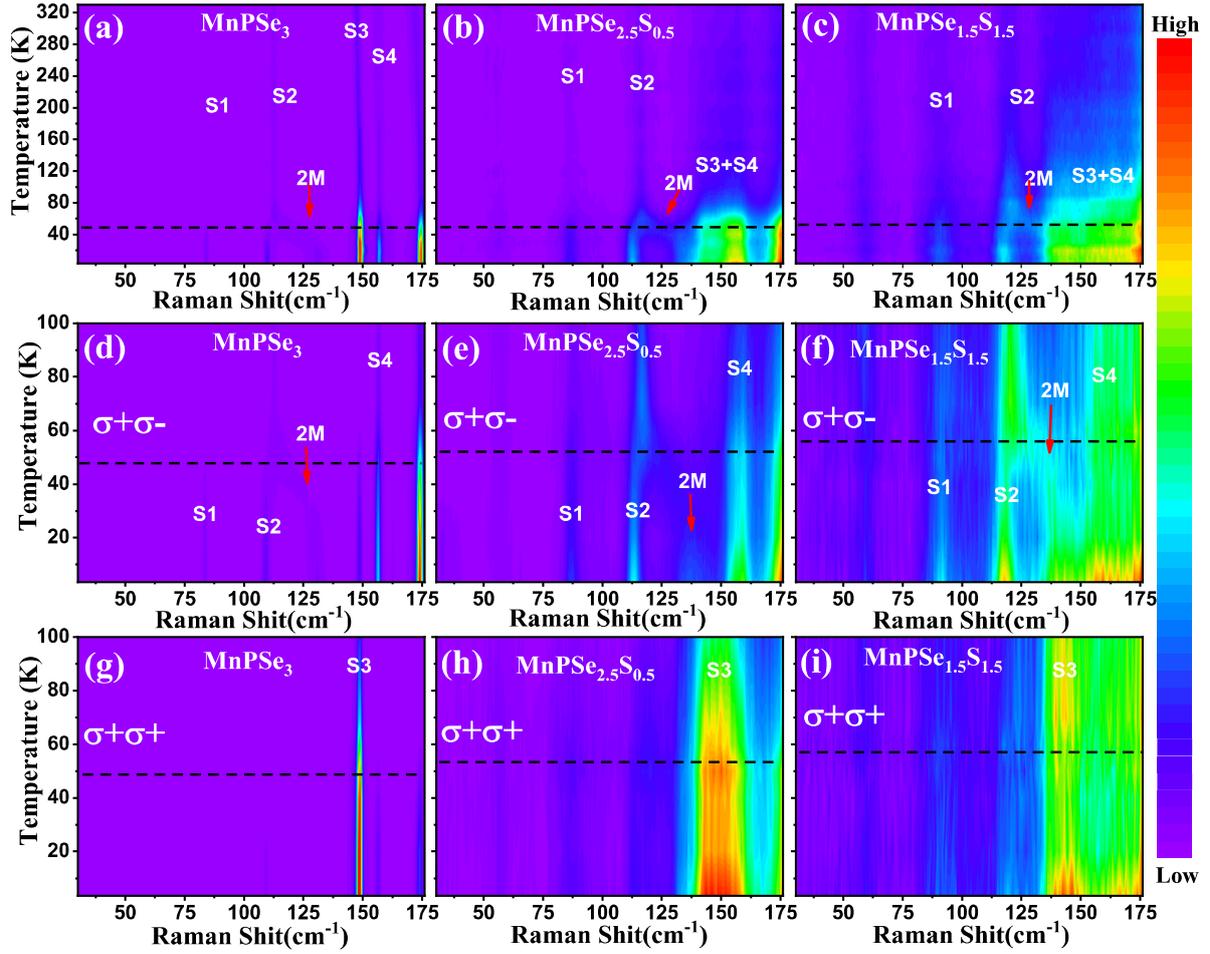

**Figure 3:** Unpolarized 2D color contour maps of the Raman intensity versus Raman shift and as a function of temperature (3.5-330 K) for (a) MnPSe$_3$, (b) MnPSe$_{2.5}$S$_{1.5}$ and (c) MnPSe$_{1.5}$S$_{1.5}$ in a frequency range of 30-175 cm$^{-1}$. 2D color contour maps of the Raman intensity versus Raman shift and as a function of temperature (3.5-100 K) for (d) MnPSe$_3$, (e) MnPSe$_{2.5}$S$_{1.5}$ and (f) MnPSe$_{1.5}$S$_{1.5}$ collected in cross-circularly ($\sigma^+\sigma^-$); (g) MnPSe$_3$, (h) MnPSe$_{2.5}$S$_{1.5}$ and (i) MnPSe$_{1.5}$S$_{1.5}$ collected in co-circularly ($\sigma^+\sigma^+$) polarized configuration in a frequency range of 30-175 cm$^{-1}$. Two-magnon (2M) signal is marked by a red arrow. The horizontal black dashed line indicates the magnon-phonon hybridization temperature.

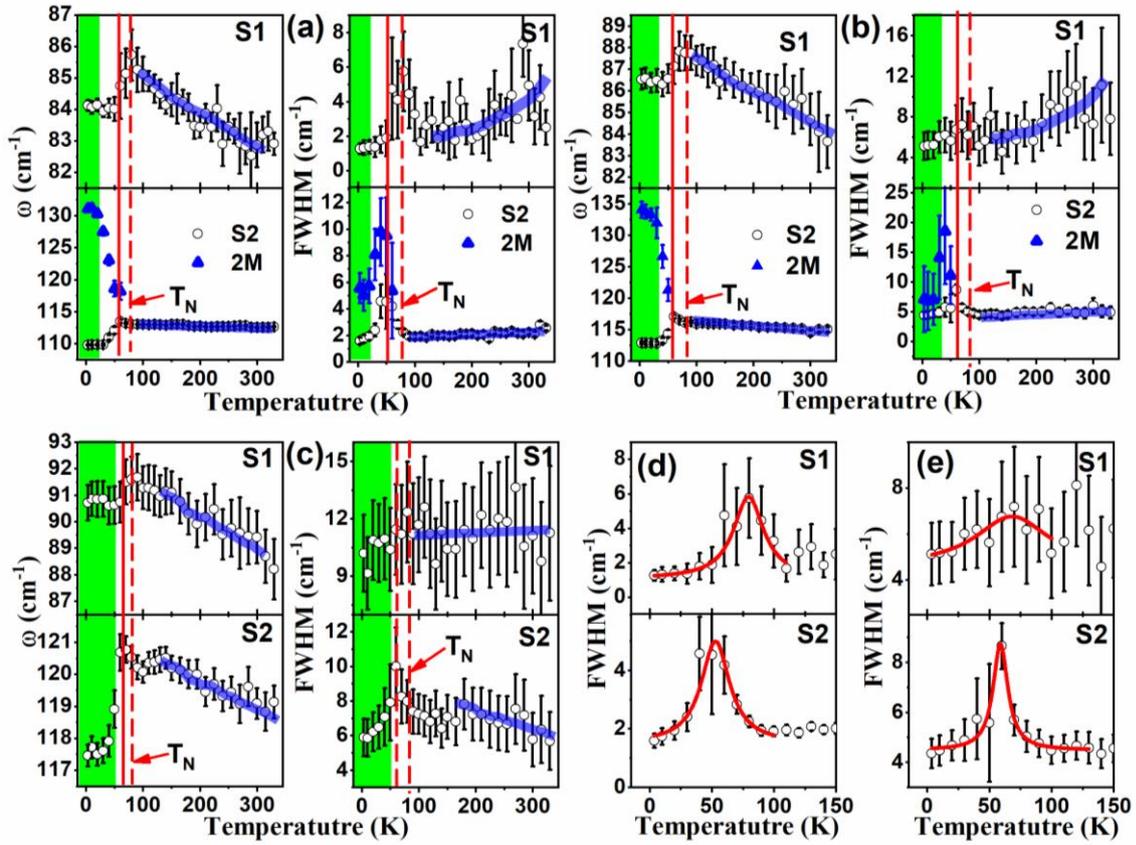

**Figure 4:** (a), (b), and (c) Temperature-dependent frequency and FWHM of S1, S2, and 2M for $MnPS_3$, $MnPSe_{2.5}S_{0.5}$, and $MnPSe_{1.5}S_{1.5}$, respectively. A solid vertical red line around 50-60 indicates the hybridization of two-magnon with S2 ($E_g$). Solid vertical red dashed line at $T_N$ indicates the hybridization of two-magnon with S1($E_g$). (d) and (e) Temperature dependent FWHM of S1 and S2 in the temperature range of 3.5 to 150K for $MnPS_3$ and $MnPSe_{2.5}S_{0.5}$, plotted for clear visibility, respectively. The green shaded area shows the topological phase region (T1). Solid red curves in (d) and (e) are the fits using the Lorentzian function. Semi-transparent blue lines are guides to the eye.

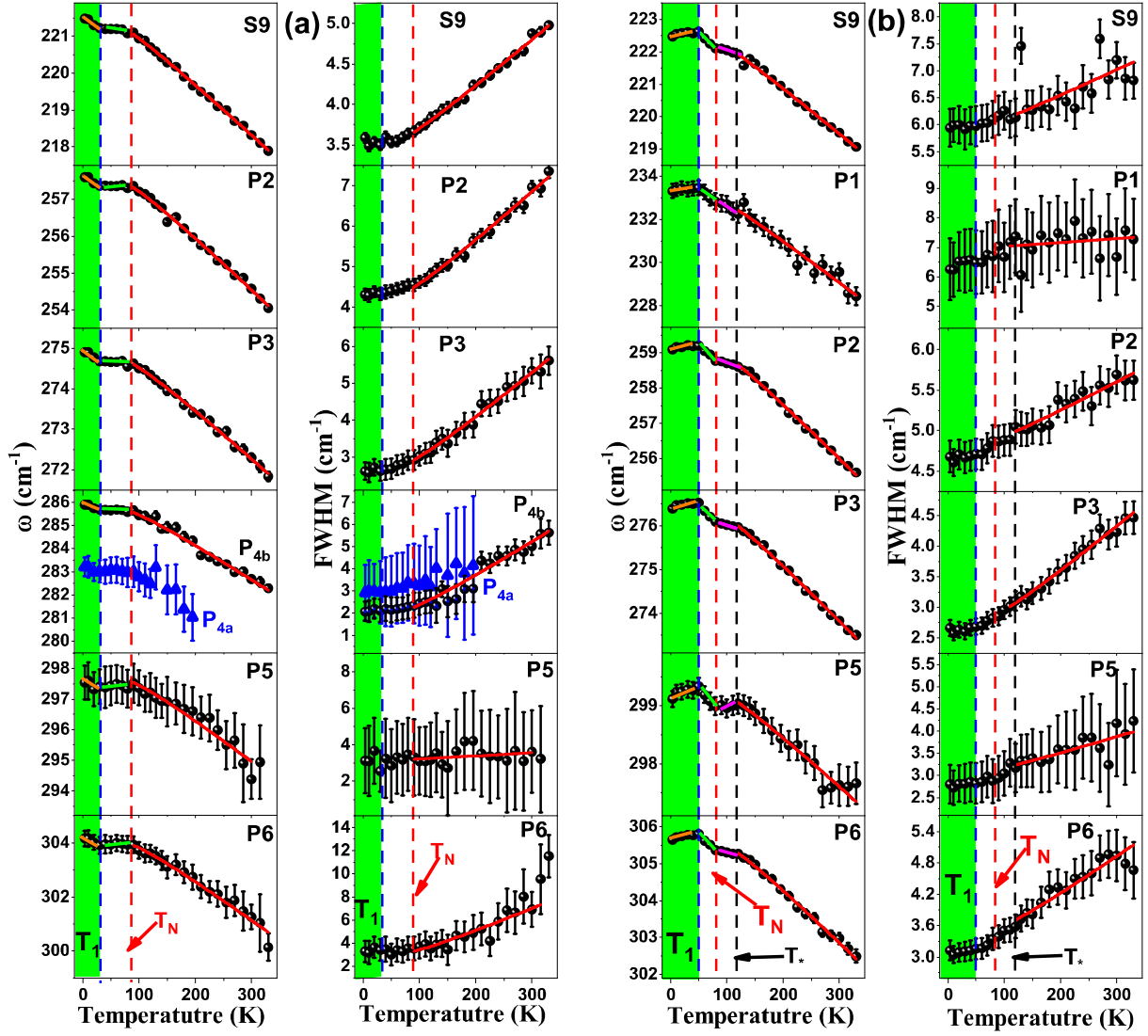

**Figure 5:** (a) Temperature-dependent frequency and FWHM of S9 and P2-P6 for MnPSe$_{2.5}$S$_{0.5}$. (b) Temperature-dependent frequency and FWHM of S9 and P1-P3, P5, and P6 for MnPSe$_{1.5}$S$_{1.5}$. The green shaded area shows the topological phase region (T1). The blue and red dashed lines correspond to the topological phase and antiferromagnetic ordering (T$_N$) transition temperature, respectively. The black dashed line for the case of MnPSe$_{1.5}$S$_{1.5}$ indicates the presence of the short-range spin-spin corrections. The solid red lines above T$_N$ for MnPSe$_{2.5}$S$_{0.5}$ and T$_*$ for MnPSe$_{1.5}$S$_{1.5}$ are the fitted curves using phonon-phonon anharmonicity interactions as described in the text. The orange, green, and magenta solid lines are guides to the eye.

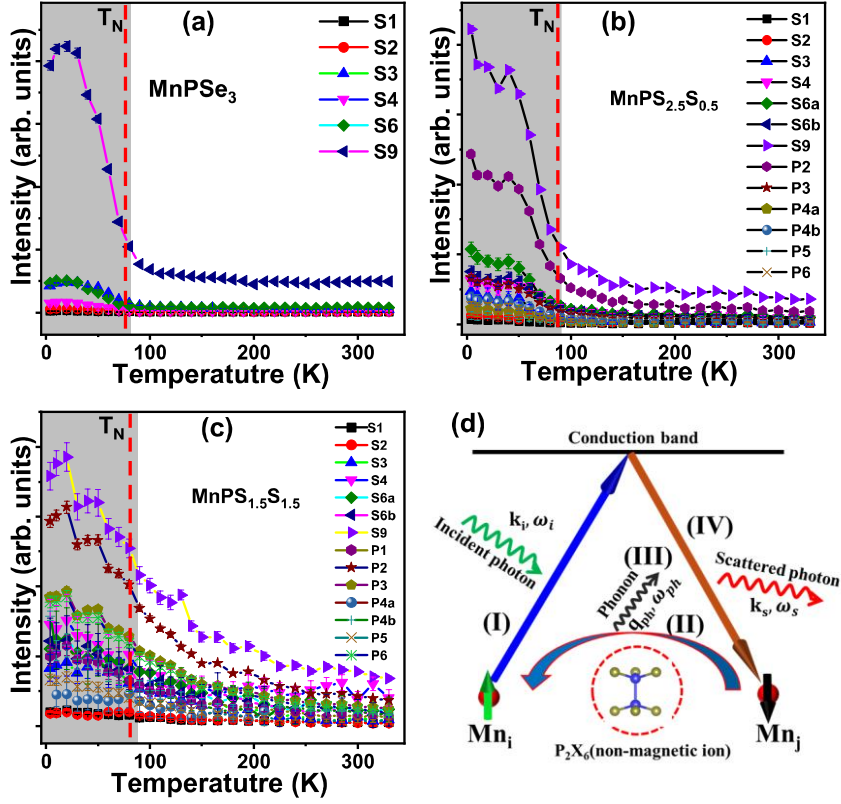

**Figure 6:** (a), (b), and (c) Temperature-dependent intensity for some of the selected phonons for MnPSe$_3$, MnPSe$_{2.5}$S$_{0.5}$, and MnPSe$_{1.5}$S$_{1.5}$, respectively. The shaded gray area shows the temperature regime where spin-dependent Raman scattering is observed. The red dashed line corresponds to the antiferromagnetic ordering (T$_N$) transition temperature. (d) Schematic representation of spin-dependent Stokes Raman scattering process.

**Supplementary Information**

# Interplay between magnetic and lattice excitations and emergent multiple phase transitions in MnPSe$_{3-x}$S$_x$

Deepu Kumar[1], Nguyen The Hoang[1], Yumin Sim[1], Youngsu Choi[2], Kalaivanan Raju[3], Rajesh Kumar Ulaganathan[3], Raman Sankar[3], Maeng-Je Seong[1]\*, and Kwang-Yong Choi[2]\*

[1]*Department of Physics and Center for Berry Curvature-based New Phenomena (BeCaP) Chung-Ang University, Seoul 06974, Republic of Korea*
[2]*Department of Physics, Sungkyunkwan University, Suwon 16419, Republic of Korea*
[3]*Institute of Physics, Academia Sinica, Taipei 10617, Taiwan*

\* Email: mseong@cau.ac.kr (M. J. Seong), choisky99@skku.edu (K.Y. Choi)

**Section S1. Crystal and Magnetic structures**

Figures S1(a) and (b) demonstrate the crystal structures of both MnPSe$_3$ and MnPS$_3$, respectively. Bulk MnPSe$_3$ crystalizes into a trigonal/rhombohedral crystal structure with a point group $C_{3i}$ and space group $R\bar{3}$ (#148). Mn atoms are arranged in a honeycomb lattice structure in the *ab* plane and each Mn atom is connected with six Se atoms in trigonal symmetry creating MnSe$_6$ octahedra. Further, Se atoms are coordinated with two P atoms above and below the Mn atoms lattice plane like a dumbbell creating the $(P_2Se_6)^{4-}$ cluster. Bulk MnPS$_3$ crystalizes into a monoclinic structure with a point group $C_{2h}$ and space group $C2/m$ (#12). Similar to MnPSe$_3$, Mn atoms in MnPS$_3$ are arranged in a honeycomb lattice in the *ab* plane and surrounded by six S atoms forming the MnS$_6$ octahedra, and S atoms are connected with two P atoms creating the $(P_2S_6)^{4-}$ cluster.

Figure S1(c) and (d) show the magnetic structures of MnPSe$_3$ and MnPS3, respectively. Both MnPSe$_3$ and MnPS$_3$ exhibit the Neel-type antiferromagnetic ordering but different spin alignments with respect to the magnetic plane. The spins are aligned parallel (perpendicular) to the magnetic plane for MnPSe$_3$ (MnPS$_3$) [1,2].

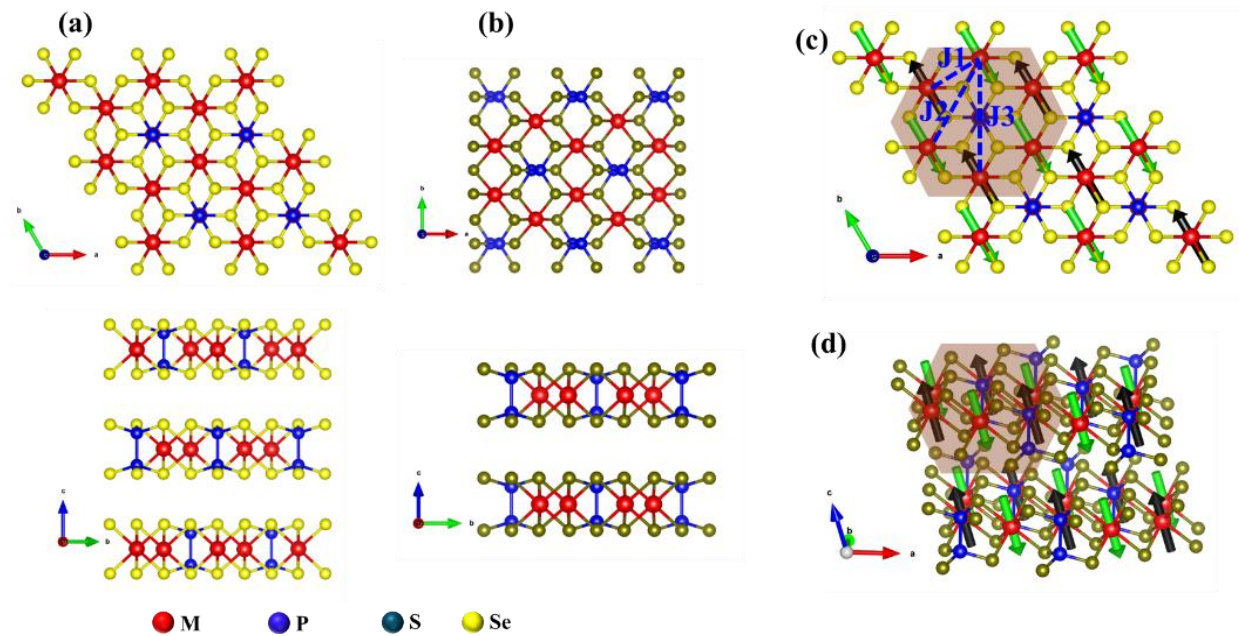

**Figure S1:** Crystal structures of (a)MnPSe$_3$ and (b) MnPS$_3$. (c) and (d) Magnetic structures of MnPSe$_3$ and MnPS$_3$, respectively. Spins are black and green color-coded to differentiate between the two spin orientations. *J1, J2*, and *J3* in (c) represent the first, second, and third nearest neighbor exchange interactions, respectively. Both crystal and magnetic structures were created using the VESTA visualization program [3].

**Section S2. Phonon excitations**

The unit cell Bulk MnPSe$_3$ is composed of two atomic formulas (Z=2) with 10 atoms (two $Mn^{2+}$ and one $(P_2Se_6)^{4-}$ cluster), which give rise to a total of 30 phonon modes at the Γ point of the Brillouin zone and are given by the following irreducible representation $\Gamma = 5A_g + 5E_g + 5A_u + 5E_u$ [4]. Among these, 15 are Raman-active ($\Gamma_{Raman} = 5A_g + 5E_g$), 12 are infrared-active ($\Gamma_{Infrared} = 4A_u + 4E_u$) and the remaining three are acoustic phonons ($\Gamma_{acoustic} = A_u + E_u$). Phonons with $E$ symmetry are doubly degenerate. Similar to MnPSe$_3$, the unit cell of MnPS$_3$ is also composed of two atomic formulas ($Z=2$) with 10 atoms, which give rise to a total of 30 phonon modes at the $\Gamma$ point of the Brillouin zone with the following irreducible representation $\Gamma = 8A_g + 7B_g + 6A_u + 9B_u$ with 15 Raman-active $\Gamma = 8A_g + 7B_g$, 12 infrared-active $\Gamma = 5A_u + 7B_u$ and the remaining 3 are acoustic phonons ($\Gamma = A_u + 2B_u$) [5,6]. Moreover, we note that although the bulk MnPS$_3$ belongs to the point group $C_{2h}$, the $(P_2S_6)^{4-}$ unit belongs to the hexagonal $D_{3d}$ symmetry which gives rise to the following irreducible representation of the phonon modes as $\Gamma = 3A_{1g} + 2A_{2g} + 5E_g + A_{1u} + 4A_{2u} + 5E_u$, where $A_{1g}$ and $E_g$ are the Raman-active while $A_{2g}$ and $A_{1u}$ modes are optically inactive [6].

The Raman tensors of the Raman-active $A_g$ and $E_g$ phonons with $C_{3i}$ point symmetry group can be expressed as

$$R(A_g) = \begin{pmatrix} a & 0 & 0 \\ 0 & a & 0 \\ 0 & 0 & b \end{pmatrix} \text{ and } R(E_g) = \begin{pmatrix} c & d & e \\ d & -c & f \\ e & f & 0 \end{pmatrix}, R(E_g) = \begin{pmatrix} d & -c & -f \\ -c & -d & e \\ -f & e & 0 \end{pmatrix}. \quad (A)$$

The Raman tensors of the Raman-active $A_{1g}$ and $E_g$ phonons with $D_{3d}$ point symmetry group are given as

$$R(A_{1g}) = \begin{pmatrix} a & 0 & 0 \\ 0 & a & 0 \\ 0 & 0 & b \end{pmatrix} \text{ and } R(E_g) = \begin{pmatrix} c & 0 & 0 \\ 0 & -c & d \\ 0 & d & 0 \end{pmatrix}, R(E_g) = \begin{pmatrix} 0 & -c & -d \\ -c & 0 & 0 \\ d & 0 & 0 \end{pmatrix}. \quad (B)$$

The Raman tensors for Raman-active $A_g$ and $B_g$ phonons with $C_{2h}$ point symmetry group can be expressed as

$$R(A_g) = \begin{pmatrix} a & 0 & d \\ 0 & b & 0 \\ d & 0 & c \end{pmatrix} \text{ and } R(B_g) = \begin{pmatrix} 0 & e & 0 \\ e & 0 & f \\ 0 & f & 0 \end{pmatrix}. \tag{C}$$

**Section S3. Selection rules for circularly polarized Raman scattering**

To shed light on the symmetry assignment of the phonons, as well as to distinguish the magnetic excitations from phonons with different symmetry, we did circularly polarized Raman measurements for all samples. The polarization direction of the circularly polarized light could be given as $\sigma^+ = [1 \ -i \ 0]/\sqrt{2}$ and $\sigma^- = [1 \ i \ 0]/\sqrt{2}$. Within the semi-classical approximation, the Raman scattering intensity of the first-order phonon modes is given as $I_{int} = |\hat{e}_s^t . R . \hat{e}_i|^2$, where $R$, $\hat{e}_i$ and $\hat{e}_s$ are the Raman tensor, the polarization vectors of the incident, and the scattered light, respectively [7,8]. The Raman scattering intensity of the $A_g / A_{1g}$ mode for the $C_{3i}$ and $D_{3d}$ point groups is given as $I_{Ag/A1g}(\sigma^+\sigma^+) \sim a^2$ and $I_{Ag/A1g}(\sigma^+\sigma^-) \sim 0$. The scattering intensity of the $E_g$ mode is given as: $I_{E_g}(\sigma^+\sigma^+) \sim 0$ and $I_{E_g}(\sigma^+\sigma^-) \sim c^2 + d^2$ for the $C_{3i}$ point group, and $I_{E_g}(\sigma^+\sigma^+) \sim 0$ and $I_{E_g}(\sigma^+\sigma^-) \sim c^2$ for the $D_{3d}$ point group. From the above selection rules, it is clear that the $A_g / A_{1g}$ and $E_g$ modes are allowed in co-circular and cross-circular polarized configurations, respectively.

The Raman scattering intensity of the $A_g$ and $B_g$ mode with $C_{2h}$ point group is given as $I_{Ag}(\sigma^+\sigma^+) \sim |a+b|^2 / 4$ and $I_{Ag}(\sigma^+\sigma^-) \sim |a-b|^2 / 4$ and $I_{Bg}(\sigma^+\sigma^+) \sim 0$ and $I_{Bg}(\sigma^+\sigma^-) \sim e^2$, respectively. In previous linearly polarized Raman studies, $a \cong b$ was obtained for $A_g$ phonon mode [9]. Considering this $a \cong b$, the selection rule allowed $A_g$ phonon mode in a co-circular polarized configuration. On the other hand, the $B_g$ mode is allowed in cross-curricular polarization configuration.

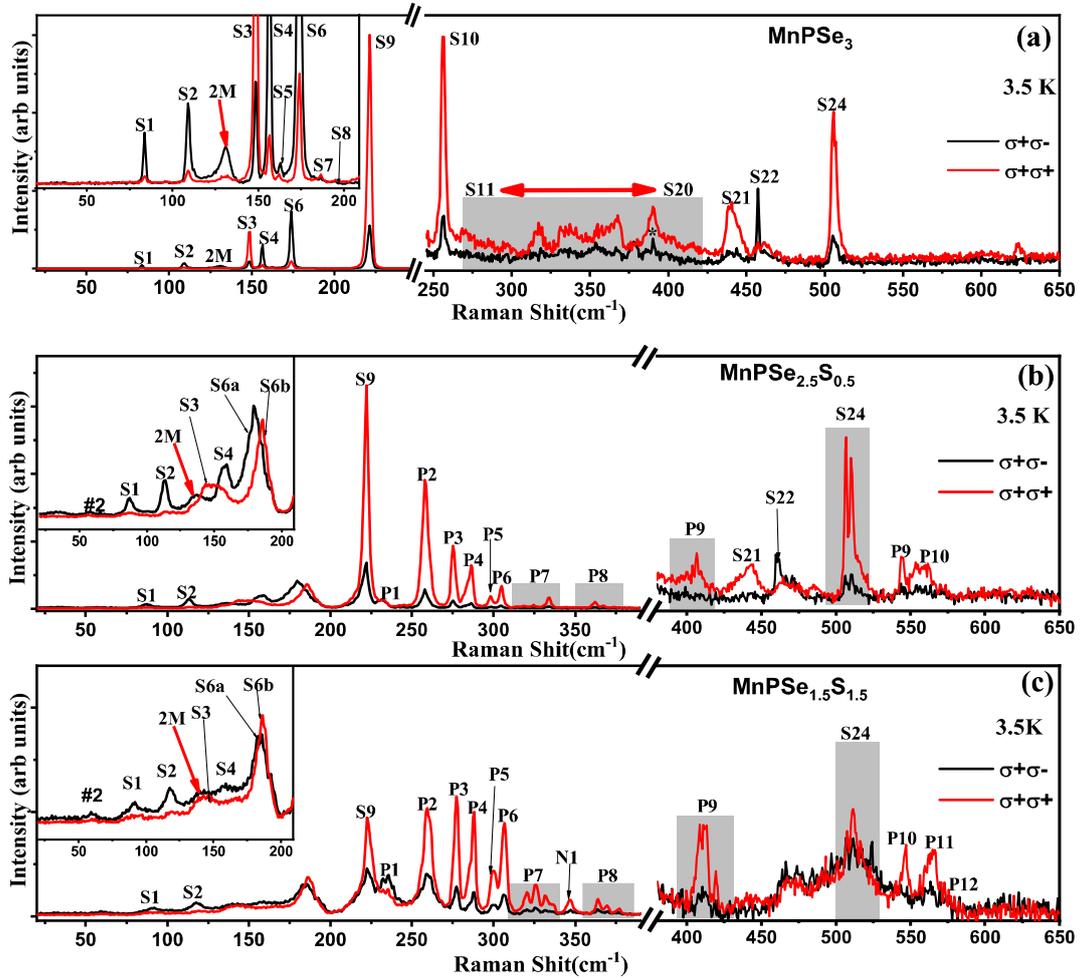

**Figure S2:** Circularly polarized Raman spectra of (a) MnPSe$_3$ (b) MnPSe$_{2.5}$S$_{0.5}$ and (c) MnPSe$_{1.5}$S$_{1.5}$ collected at 3.5 K in co-circularly ($\sigma^+\sigma^+$; red) and cross-circularly ($\sigma^+\sigma^-$; black) configurations. Two-magnon is marked by a red arrow.

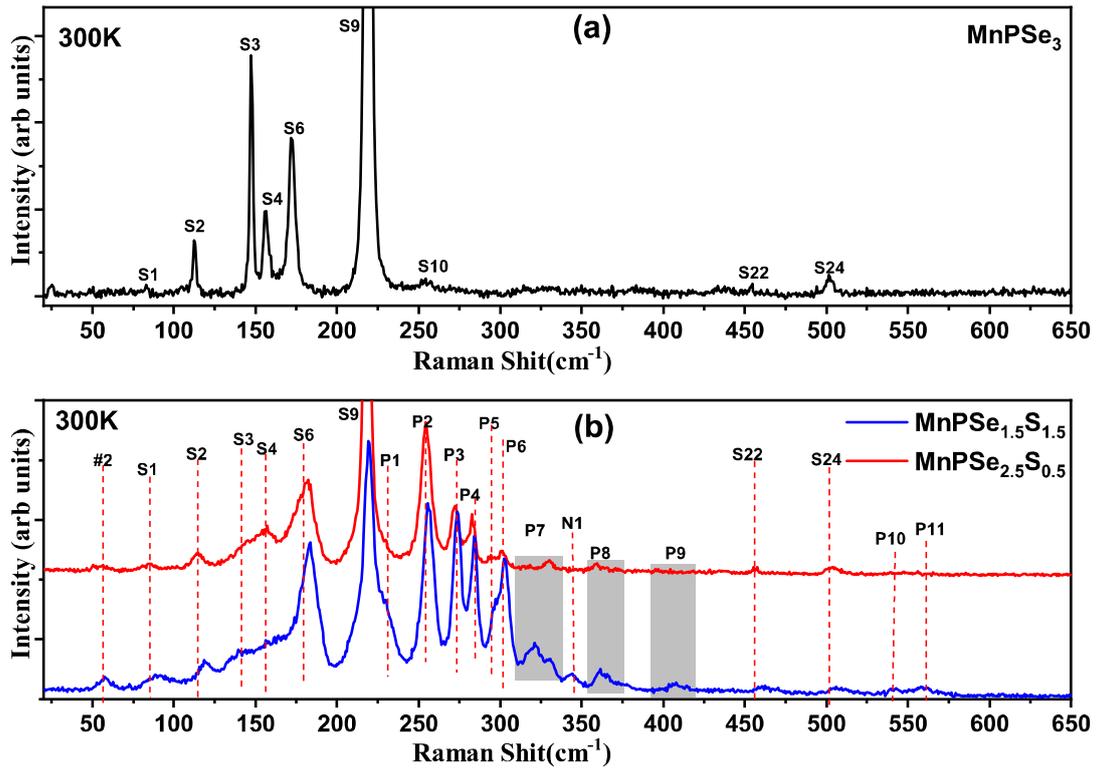

**Figure S3:** Unpolarized Raman spectrum of (a) $MnPSe_3$ and (b) $MnPSe_{2.5}S_{0.5}$ (red); $MnPSe_{1.5}S_{1.5}$ (blue) collected at 300 K.

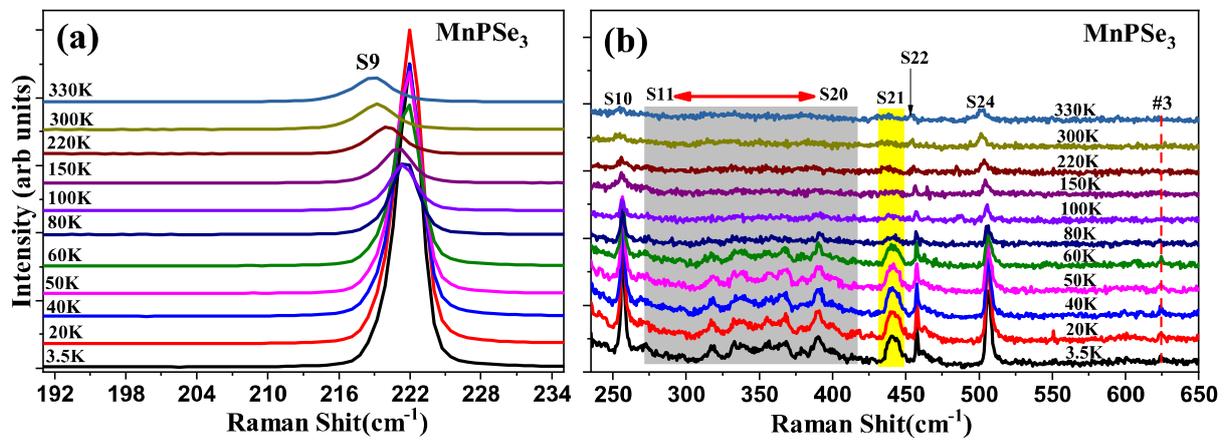

**Figure S4:** (a) and (b) Temperature evolution of the unpolarized Raman spectrum of $MnPSe_3$ in the frequency range of 190-235 $cm^{-1}$ and 230-650 $cm^{-1}$, respectively.

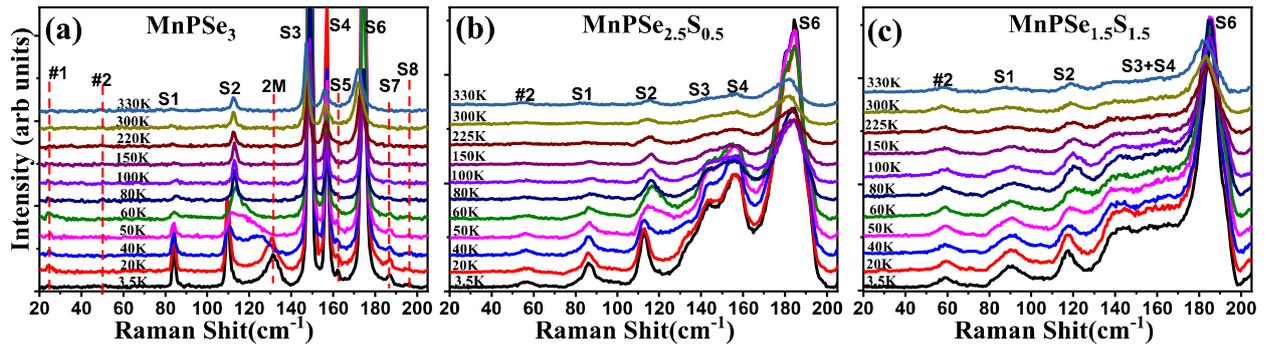

**Figure S5:** Temperature evolution of the unpolarized Raman spectrum of (a) MnPSe$_3$, (b) MnPSe$_{2.5}$S$_{0.5}$, and (c) MnPSe$_{1.5}$S$_{1.5}$ in a frequency range of 20-200 cm$^{-1}$.

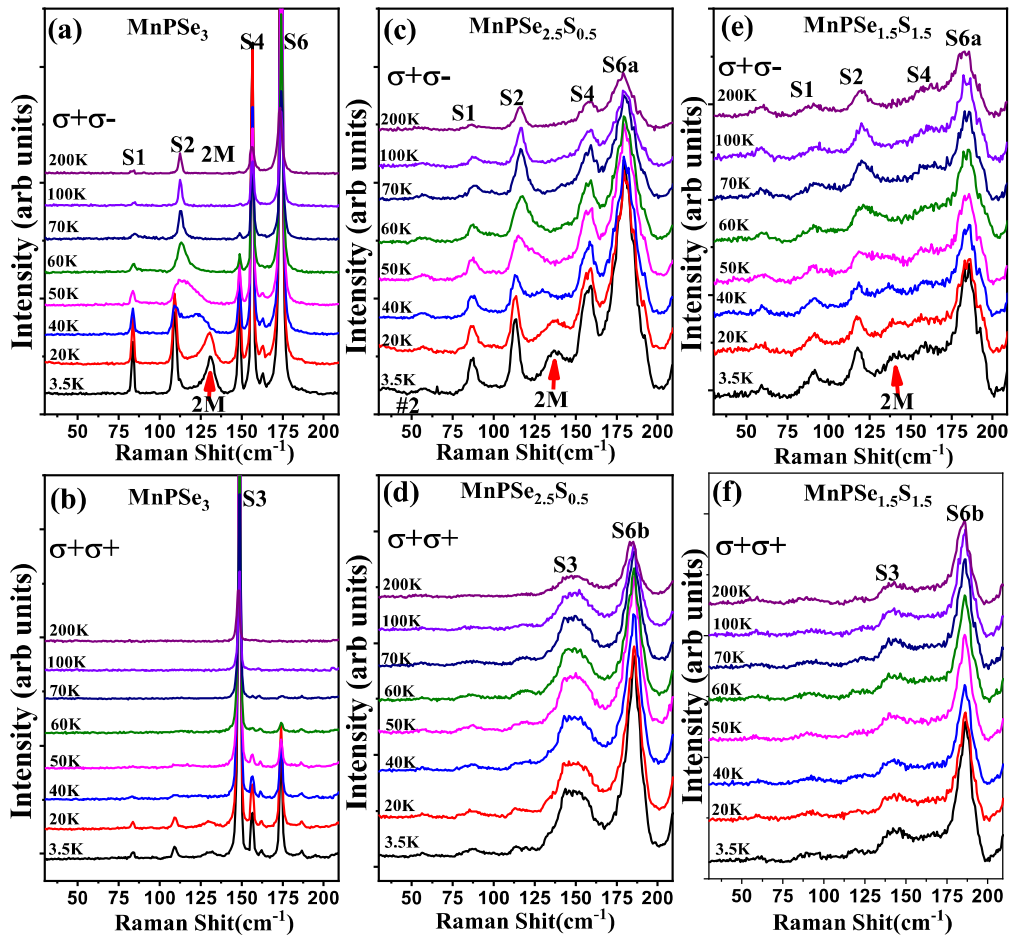

**Figure S6:** Temperature evolution of the Raman spectrum of (a) MnPSe$_3$, (c) MnPSe$_{2.5}$S$_{0.5}$ and (d) MnPSe$_{1.5}$S$_{1.5}$ collected in cross-circularly ($\sigma^+\sigma^-$) ; (b) MnPSe$_3$, (d) MnPSe$_{2.5}$S$_{0.5}$ and (f) MnPSe$_{1.5}$S$_{1.5}$ collected in co-circularly ($\sigma^+\sigma^+$) configurations in a frequency range of 30-200 cm$^{-1}$. Two-magnon signal is marked by a red arrow.

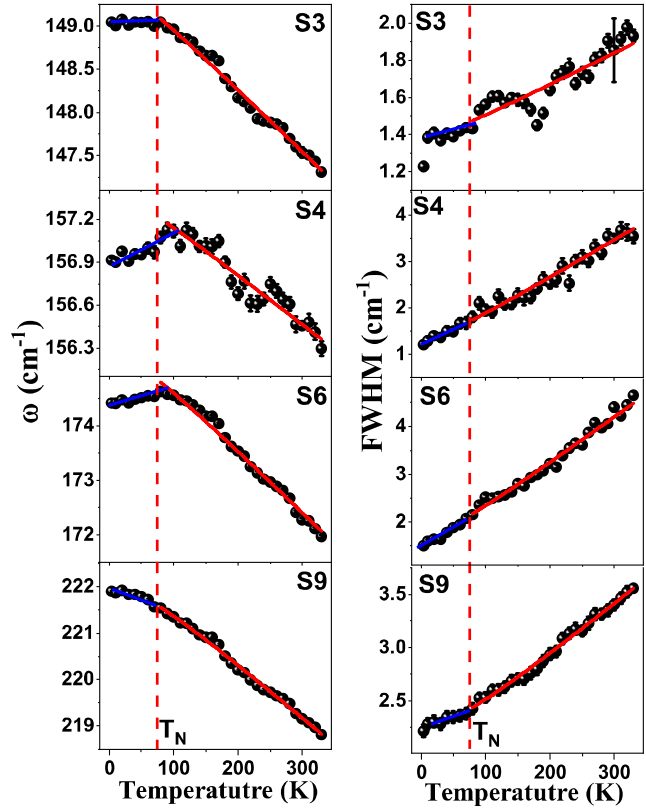

**Figure S7:** Temperature-dependent frequency and FWHM of S3, S4, S6 and S9 for MnPS$_3$. The red dashed line corresponds to T$_N$. The solid red lines are fitted curves using three phonons anharmonicity as described in the main text. Solid blue lines are a guide to the eye.

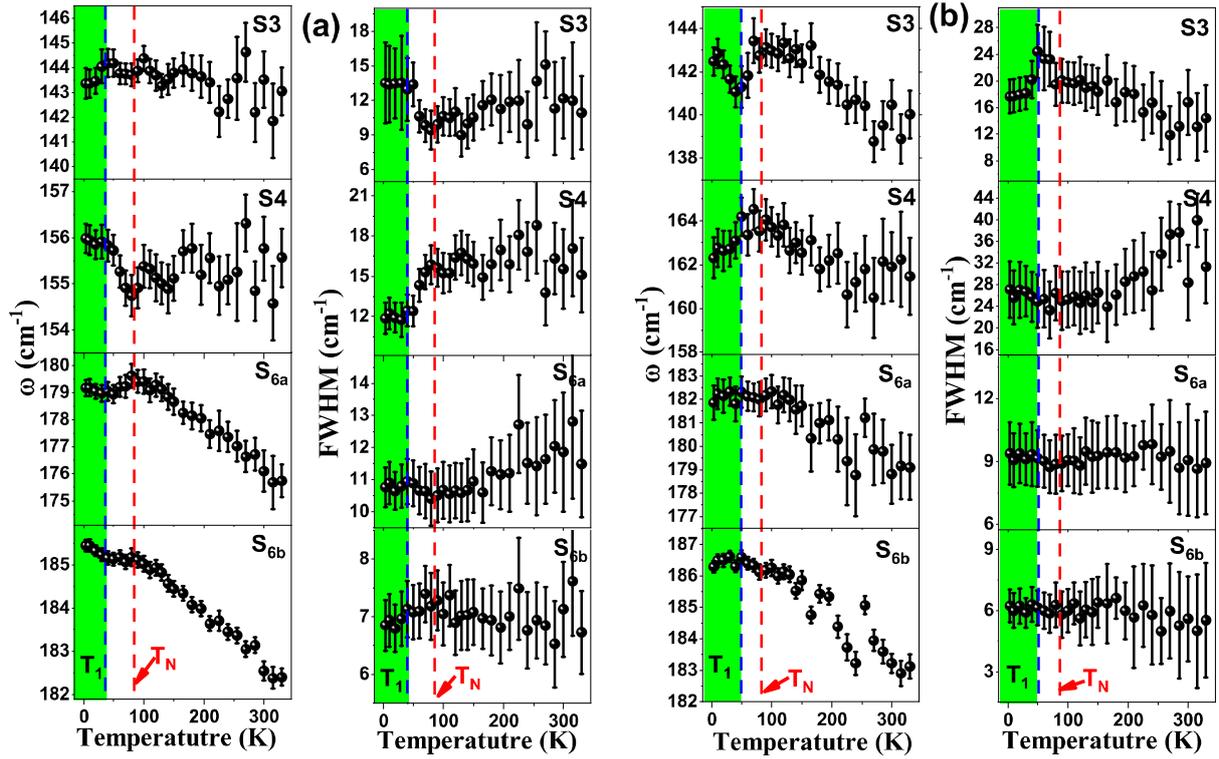

**Figure S8:** (a) and (b)Temperature-dependent frequency and FWHM of S3, S4, and S6 for MnPSe$_{2.5}$S$_{0.5}$ and MnPSe$_{1.5}$S$_{1.5}$, respectively. The green shaded area shows the topological phase region (T1). The blue and red dashed lines correspond to the topological phase and antiferromagnetic ordering (T$_N$) transition temperature, respectively.

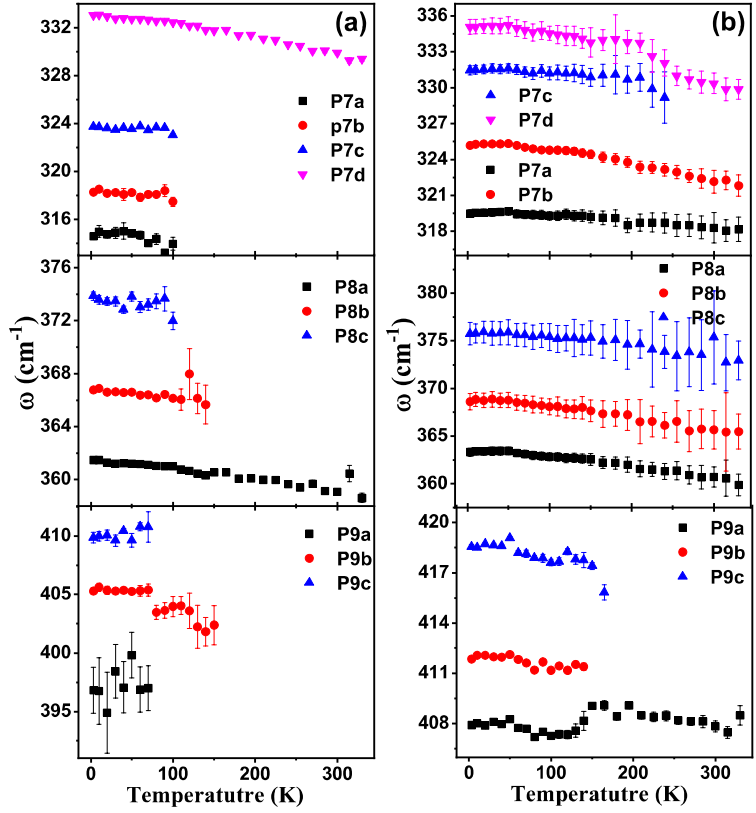

**Figure S9:** Temperature-dependent frequency of the P7, P8, and P9 grouped phonons.

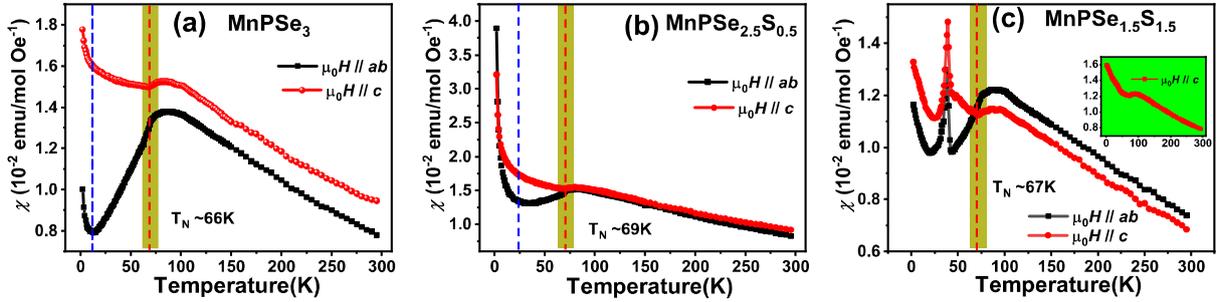

**Figure S10:** Temperature dependence of magnetic susceptibility for (a) MnPSe$_3$ (b) MnPSe$_{1.5}$S$_{1.5}$ and (c) MnPSe$_{1.5}$S$_{1.5}$ measured with parallel (black) and perpendicular (red) to the *ab* plane at an external field of $\mu_0 H = 0.02\,\text{T}$. Inset (c) is the temperature-dependent magnetic susceptibility for MnPSe$_{1.5}$S$_{1.5}$ at an external field of $\mu_0 H = 1\,\text{T}$. The blue and red dashed lines correspond to the topological phase and antiferromagnetic ordering ($T_\text{N}$) transition temperature, respectively.

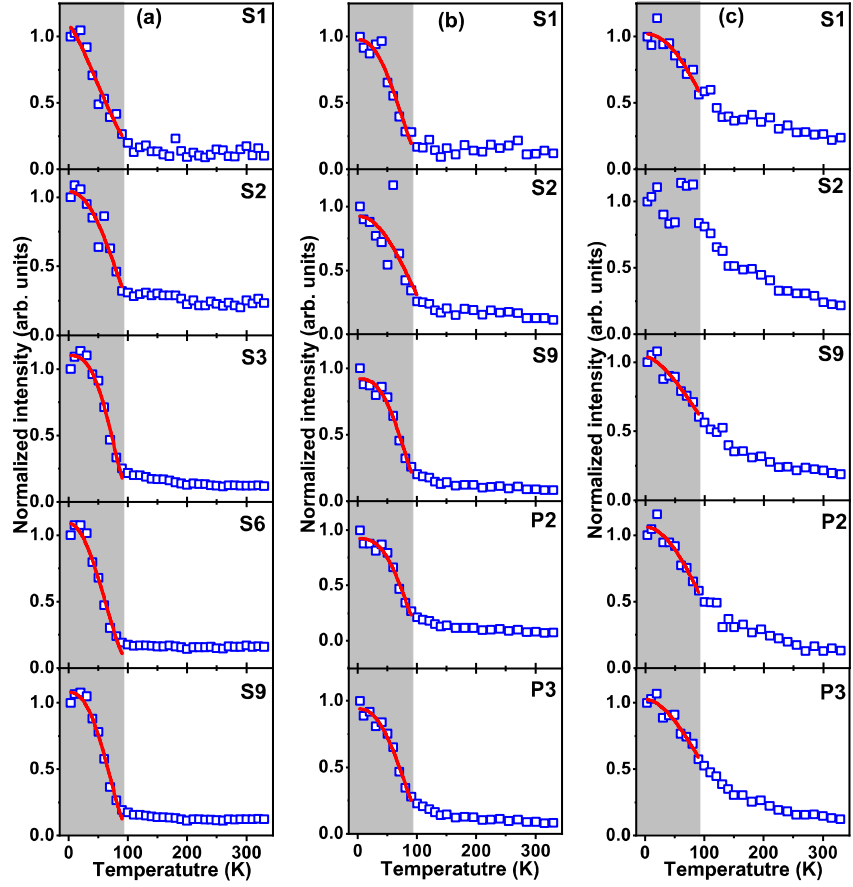

**Figure S11:** (a) Temperature-dependent normalized intensity of S1, S2, S3, S6, and S9 modes for MnPSe$_3$. (a) and (b) Temperature-dependent intensity of S1, S2, S9, P2, and P3 for MnPSe$_{2.5}$S$_{0.5}$ and MnPSe$_{1.5}$S$_{1.5}$, respectively. The gray shaded area shows the temperature range where spin-dependent Raman scattering is observed. Solid red lines are fits as described in the main text.

**Table-S1**: List of the experimentally observed modes along with their symmetry assignment and frequency at 3.5 K for MnPS$_{3-x}$S$_x$ ($x$=0, 0.5, 1.5). Units are in cm$^{-1}$

| Mode | Frequency at 3.5K | | | |
|---|---|---|---|---|
| | MnPSe$_3$ | MnPSe$_{2.5}$S$_{0.5}$ | MnPSe$_{1.5}$S$_{0.5}$ | |
| #1 | ~24 | - | - | |
| #2 | ~50 | 56.5 | 59.1 | |
| S1 | 84.1 | 86.5 | 90.7 | Eg_Mn$^{2+}$ |
| S2 | 109.8 | 112.8 | 117.5 | Eg_ Mn$^{2+}$ |
| 2M | 131.1 | 134.0 | | |
| S3 | 149.0 | 143.4 | 143.7 | Ag _P$_2$Se$_6$ |
| S4 | 156.9 | 156.0 | 163.4 | Eg _P$_2$Se$_6$ |
| S5 | 162.2 | - | - | Ag |
| S6 | 174.4 | a-179.2<br>b-185.4 | 181.4<br>186.1 | a-Eg<br>b-Ag |
| S7 | 186.7 | - | - | Ag |
| S8 | 195.6 | - | - | Ag |
| S9 | 221.9 | 221.5 | 222.5 | Ag _P$_2$Se$_6$ |
| P1 | | 232.0 | 233.3 | Bg/Ag _P$_2$S$_6$ |
| P2 | | 257.6 | 259.1 | Ag _P$_2$S$_6$ |
| S10 | 256.7 | - | - | Ag |
| S11 | 272.2 | - | - | Ag |
| P3 | | 274.9 | 276.4 | Ag/Bg_P$_2$S$_6$ |
| P4 | | a-283.3<br>b-285.9 | 285.2<br>287.2 | Ag |
| S12 | 292.9 | | | Ag |
| P5 | | 297.6 | 299.1 | Ag |
| P6 | | 304.1 | 305.6 | Ag |
| S13 | 317.7 | | | Ag |
| P7 | | a-314.6<br>b-318.3<br>c-323.7<br>d-333.1 | 319.5<br>325.2<br>331.5<br>335.1 | Ag |
| S14 | 332.5 | | | Ag |
| S15 | 338.8 | | | Ag |
| S16 | 355.9 | | | Ag |
| S17 | 367.5 | | | Ag |
| N1 | | | 346.1 | Ag |
| P8 | | a-361.5<br>b-366.7<br>c-373.9 | 363.3<br>368.6<br>375.7 | Ag/Bg (P$_2$S$_6$) |
| S18 | 379.1 | | | Ag |
| S19 | 389.9 | | | Ag |
| S20 | 399.3 | | | Ag |
| P9 | | a-396.9<br>b-405.3<br>c-409.9 | 407.9<br>411.9<br>418.5 | Ag |
| S21 | 441.5 | a-438.8<br>b-443.8 | 442.7 | Ag |
| S22 | 457.8 | 460.1 | 464.3 | Eg |
| S23 | 462.7 | 467.6 | 470.9 | |
| S24 | 506.3 | a-505.9<br>b-509.8<br>c-515.3 | 506.6<br>510.9<br>516.5 | Ag |
| P10 | | 543.7 | 545.2 | Ag |
| P11 | | 555.4 | 563.3 | Ag |
| P12 | | | 573.4 | Ag |

**Table-S2:** List of the extracted approximate phase transitions temperature using the Raman and magnetic susceptibility ($\chi(\mu_0 H // ab)$).

| Phase transition extracted from Raman [ K] | | | | Phase transition extracted from magnetic susceptibility ($\chi(\mu_0 H // ab)$) [ K] | | |
|---|---|---|---|---|---|---|
| Phase | MnPSe$_3$ | MnPSe$_{2.5}$S$_{0.5}$ | MnPSe$_{1.5}$S$_{1.5}$ | MnPSe$_3$ | MnPSe$_{2.5}$S$_{0.5}$ | MnPSe$_{1.5}$S$_{1.5}$ |
| Topological (T1) | 17.5 | 33 | 50 | 12 | 23 | - |
| Long-ranged ordering (T$_N$) | 76 | 85 | 80.8 | 66 | 69 | 67 |
| Short-ranged ordering (T*) | 85 | 90 | 117.3 | 81 | 82 | 90 |

**Table-S3**: List of best-fit parameters, for some of the selected phonon modes, using spin-dependent scattering expression as described in the main text.

| | Mode | R | M | $\gamma$ |
|---|---|---|---|---|
| **MnPSe$_3$** | S1 | 0.61±0.04 | -0.43±0.05 | 1.34±0.2 |
| | S2 | 0.72±0.03 | -0.30±0.04 | 2.15±0.4 |
| | S3 | 0.66±0.02 | -0.39±0.03 | 2.78±0.5 |
| | S6 | 0.53±0.03 | -0.51±0.04 | 1.99±0.3 |
| | S9 | 0.58±0.03 | -0.45±0.03 | 2.39±0.3 |
| | | | | |
| **MnPSe$_{2.5}$S$_{0.5}$** | S1 | 0.51±0.05 | -0.48±0.06 | 2.2±0.5 |
| | S2 | 0.67±0.07 | -0.29±0.1 | 2.08±0.4 |
| | S9 | 0.54±0.04 | -0.42±0.04 | 2.63±0.5 |
| | P1 | 0.55±0.03 | -0.41±0.04 | 2.74±0.5 |
| | P2 | 0.56±0.03 | -0.41±0.03 | 2.33±0.4 |
| | | | | |
| **MnPSe$_{1.5}$S$_{1.5}$** | S1 | 0.82±0.02 | -0.19±0.03 | 2.12±0.8 |
| | S2 | - | - | - |
| | S9 | 0.83±0.02 | -0.19±0.02 | 1.59±0.5 |
| | P1 | 0.81±0.02 | -0.22±0.03 | 1.98±0.6 |
| | P2 | 0.82±0.02 | -0.20±0.02 | 1.81±0.5 |

**References:**

[1]  S. Calder, A. V. Haglund, A. I. Kolesnikov, and D. Mandrus, Magnetic Exchange Interactions in the van Der Waals Layered Antiferromagnet Mn PSe$_3$, Phys. Rev. B **103**, 024414 (2021).

[2]  R. Basnet, K. M. Kotur, M. Rybak, C. Stephenson, S. Bishop, C. Autieri, M. Birowska, and J. Hu, Controlling Magnetic Exchange and Anisotropy by Nonmagnetic Ligand Substitution in Layered M PX$_3$ (M=Ni, Mn; X= S, Se), Phys. Rev. Res. **4**, 023256 (2022).

[3]  K. Momma and F. Izumi, VESTA 3 for Three-Dimensional Visualization of Crystal, Volumetric and Morphology Data, J. Appl. Crystallogr. **44**, 1272 (2011).

[4]  M. Chisa, S. Tomoyuki, T. Yoshiko, and K. Koh, Raman scattering in the two-dimensional antiferromagnet MnPSe$_3$, J. Phys Condens. Matter **5**, 623 (1993).

[5]  S. Yan, W. Wang, C. Wang, L. Chen, X. Ai, Q. Xie, and G. Cheng, Anharmonic Phonon Scattering Study in MnPS$_3$ crystal by Raman Spectroscopy, Appl. Phys. Lett. **121**, 032203 (2022).

[6]  A. Hashemi, H. P. Komsa, M. Puska, and A. V. Krasheninnikov, Vibrational Properties of Metal Phosphorus Trichalcogenides from First-Principles Calculations, J. Phys. Chem. C **121**, 27207 (2017).

[7]  W. Hayes and R. Loudon, Scattering of Light by Crystals, (John Wiley and Sons, New York, 1978).

[8]  M. Cardona and G. Guntherodt, Light Scattering in Solid II (Springer-Verlag Berlin Heidelberg New York, 1982).

[9]  K. Kim, S. Y. Lim, J. Kim, J.U. Lee, S. Lee, P. Kim, K. Park, S. Son, C. H Park, J. G. Park, Antiferromagnetic Ordering in van Der Waals 2D Magnetic Material MnPS$_3$ Probed by Raman Spectroscopy, 2D Mater. **6**, 041001 (2019).